\documentclass[10pt,conference]{IEEEtran}

\def\BibTeX{{\rm B\kern-.05em{\sc i\kern-.025em b}\kern-.08em
    T\kern-.1667em\lower.7ex\hbox{E}\kern-.125emX}}

\usepackage[T1]{fontenc}    
\usepackage{url}            
\usepackage{booktabs}       
\usepackage{amsfonts}       
\usepackage{nicefrac}       
\usepackage{microtype}      
\usepackage{multirow}
\usepackage{enumitem}
\usepackage{amsmath,amsfonts}
\usepackage{bm}
\usepackage{graphicx}
\usepackage{caption}
\usepackage{listings}
\usepackage{textcomp}
\usepackage{xcolor}
\usepackage{colortbl}
\usepackage{subcaption}
\usepackage{tcolorbox}
\usepackage{xspace}
\usepackage{flushend}
\usepackage{pifont}
\usepackage{wrapfig}
\usepackage{mdframed}
\usepackage{paracol}
\usepackage{needspace}
\usepackage{caption}
\usepackage{subcaption}
\usepackage{hyperref} 

\usepackage{cite}

\makeatletter
\DeclareRobustCommand\onedot{\futurelet\@let@token\@onedot}
\def\@onedot{\ifx\@let@token.\else.\null\fi\xspace}
\def\etal{{et al}\onedot}
\makeatother

\newcommand{\rqone}{What kinds of code generation errors do different LLMs make?}

\newcommand{\rqfour}{How much effort is needed to fix code generation errors?}

\newcommand{\rqfive}{How does the task complexity affect an LLM's code generation?}

\newcommand{\rqseven}{Does partially failed code exhibit different characteristics compared with fully failed code?}

\newcommand{\code}[1]{{{\small \ttfamily #1}}}

\newcommand{\midsepremove}{\aboverulesep = 0mm \belowrulesep = 0mm}
    \midsepremove
    \newcommand{\midsepdefault}{\aboverulesep = 0mm \belowrulesep = 0mm}
    \midsepdefault

\newbool{responseHighlight}
\setbool{responseHighlight}{false}

\definecolor{darkBlue}{rgb}{0.000000,0.000000,0.545098}
\definecolor{darkGreen}{rgb}{0.000000,0.392157,0.000000}
\definecolor{DarkGray}{gray}{0.4}
\definecolor{javared}{rgb}{0.6,0,0} 
\definecolor{javagreen}{rgb}{0.25,0.5,0.35} 
\definecolor{javapurple}{rgb}{0.5,0,0.35} 
\definecolor{javadocblue}{rgb}{0.25,0.35,0.75} 
\definecolor{lightgray}{gray}{0.7}
\definecolor{lightblue}{rgb}{0.63, 0.79, 0.95}

\definecolor{shadecolor}{RGB}{150,150,150}
\definecolor{blueA}{RGB}{204,229,255}
\definecolor{redA}{RGB}{112,0, 0}
\definecolor{RED}{RGB}{255,0, 0}
\definecolor{lightred}{HTML}{FFEBEE}

\ifbool{responseHighlight}{

\newcommand{\responseref}[0]{\color{blue}}
\newcommand{\responseline}[1]{\textcolor{blue}{#1}}

\newcommand{\responsecodecommentref}[0]{\color{blue}}

\newcommand{\responsecodeoperationref}[0]{\color{blue}\bfseries}

\newcommand{\responsecodemainref}[0]{\color{blue}}

\newcommand{\responsecodestringref}[0]{\color{blue}}

}{

\newcommand{\responseref}[0]{\color{black}}
\newcommand{\responseline}[1]{\textcolor{black}{#1}}

\newcommand{\responsecodecommentref}[0]{\color{javagreen}}

\newcommand{\responsecodeoperationref}[0]{\color{javapurple}\bfseries}

\newcommand{\responsecodemainref}[0]{\color{black}}

\newcommand{\responsecodestringref}[0]{\color{javared}}

}

\newcounter{finding}
\newenvironment{finding}
{
    \refstepcounter{finding}
	\begin{mdframed}[
    	nobreak=true,
    	linecolor=black,
    	roundcorner=12pt,
    	backgroundcolor=gray!05,
    	linewidth=0.5pt,
    	leftmargin=1pt,
    	rightmargin=1pt,
        innerleftmargin=5pt,
        innerrightmargin=5pt,
    	topline=true,
    	bottomline=true,
    	skipabove=10pt
	]
    \textbf{Finding \arabic{finding}}: 
}
{
    \end{mdframed}
    \vspace{3pt}
}

\edef\oldtt{\ttdefault}
\usepackage[scaled]{beramono}
\usepackage[T1]{fontenc}
\renewcommand{\ttdefault}{\oldtt}

\lstdefinestyle{mystyle}{
  frame=single,
  xleftmargin=4pt,
  xrightmargin=4pt,
  abovecaptionskip=2pt,
  belowcaptionskip=0pt,
  captionpos=b,
  escapeinside={*‘}{’*},
  tabsize=4,
  emphstyle={\bf},
  basicstyle=\linespread{0.9}\scriptsize\ttfamily,
  keywordstyle=\color{javapurple}\bfseries,
  stringstyle=\color{javared},
  commentstyle=\color{javagreen},
  morecomment=[s][\color{javadocblue}]{/**}{*/},
  showspaces=false,
  columns=flexible,
  showstringspaces=false,
  morecomment=[l]{//},
  breaklines=true
}

\lstdefinestyle{tblstyle}{
  frame=none,
  xleftmargin=-10pt,
  xrightmargin=0pt,
  abovecaptionskip=0pt,
  belowcaptionskip=0pt,
  captionpos=b,
  escapeinside={*‘}{’*},
  tabsize=4,
  emphstyle={\bf},
  basicstyle=\linespread{0.5}\fontsize{6}{8}\ttfamily,
  keywordstyle=\color{javapurple}\bfseries,
  stringstyle=\color{javared},
  commentstyle=\color{javagreen},
  morecomment=[s][\color{javadocblue}]{/**}{*/},
  showspaces=false,
  columns=flexible,
  showstringspaces=false,
  morecomment=[l]{//},
  breaklines=true
}

\newcommand{\smalltitle}[1]{\vspace{0mm}{\noindent\bf \textit{#1}}}

\newcommand{\codeoperation}[1]{\textcolor{javapurple}{\bfseries #1}}

\newcommand{\codestring}[1]{\textcolor{javared}{#1}}

\newcommand{\codecomment}[1]{\textcolor{javagreen}{#1}}

\makeatletter
\newcommand\footnoteref[1]{\protected@xdef\@thefnmark{\ref{#1}}\@footnotemark}
\makeatother

\IEEEoverridecommandlockouts

\begin{document}

\title{Towards Understanding the Characteristics of Code Generation Errors Made by Large Language Models
}

\author{
\IEEEauthorblockN{Zhijie Wang\IEEEauthorrefmark{1}\IEEEauthorrefmark{2}, Zijie Zhou\IEEEauthorrefmark{3}\IEEEauthorrefmark{2}, Da Song\IEEEauthorrefmark{1}\IEEEauthorrefmark{2}\thanks{\IEEEauthorrefmark{2}~The first three authors contributed equally to this work. Zijie Zhou was a remote research intern at Purdue University.}, Yuheng Huang\IEEEauthorrefmark{4}, Shengmai Chen\IEEEauthorrefmark{6}, Lei Ma\IEEEauthorrefmark{4}\IEEEauthorrefmark{1}, Tianyi Zhang\IEEEauthorrefmark{6}}
\IEEEauthorblockA{
    \IEEEauthorrefmark{1}
    University of Alberta, Edmonton, AB, Canada
    \IEEEauthorrefmark{3}
    University of Illinois Urbana-Champaign, Champaign, IL, USA\\
    \IEEEauthorrefmark{4}
    The University of Tokyo, Tokyo, Japan
    \IEEEauthorrefmark{6}
    Purdue University, West Lafayette, IN, USA\\
zhijie.wang@ualberta.ca, zijiez4@illinois.edu, dsong4@ualberta.ca, yuhenghuang42@g.ecc.u-tokyo.ac.jp,\\chen3301@purdue.edu, ma.lei@acm.org, tianyi@purdue.edu}
}

\maketitle

\begin{abstract}
  Large Language Models (LLMs) have demonstrated unprecedented capabilities in code generation. However, there remains a limited understanding of code generation errors that LLMs can produce.
  To bridge the gap, we conducted an in-depth analysis of code generation errors across six representative LLMs on the HumanEval dataset. Specifically, we first employed open coding and thematic analysis to distill a comprehensive taxonomy of code generation errors. We analyzed two dimensions of error characteristics---{\em semantic characteristics} and {\em syntactic characteristics}. Our analysis revealed that LLMs often made non-trivial, multi-line code generation errors in various locations and with various root causes. We further analyzed the correlation between these errors and task complexity as well as test pass rate. 
  Our findings highlighted several challenges in locating and fixing code generation errors made by LLMs. In the end, we discussed several future directions to address these challenges.
\end{abstract}



\section{Introduction}
Automatically generating code from natural language has been a long-term pursuit across multiple research communities. 
Recent advances in Large Language Models (LLMs) have led to rapid, unprecedented improvements on this task~\cite{nijkamp2022codegen, fried2022incoder, li2023starcoder, openai2023gpt4, chatgpt2023}. 
Despite this great progress, LLMs still cannot reliably generate correct code for many tasks. Currently, there is a lack of deep understanding of the cases where LLMs fail. Specifically, it remains unclear {\em what types of code generation errors an LLM typically produces} and {\em whether different LLMs make similar errors}.
Answering these questions would help researchers gain insights into the limitations of existing models and identify opportunities for model improvement.

To bridge this knowledge gap, we conducted an in-depth analysis of code generation errors made by  LLMs. We focused on six popular LLMs: CodeGen-16B~\cite{nijkamp2022codegen}, InCoder-1.3B~\cite{fried2022incoder}, GPT-3.5~\cite{chatgpt2023}, GPT-4~\cite{openai2023gpt4}, SantaCoder~\cite{allal2023santacoder}, and StarCoder~\cite{li2023starcoder}. 
These models produced 557 incorrect code solutions on the 164 tasks from the HumanEval dataset~\cite{chen2021evaluating}. Four of the authors worked together to locate the erroneous parts of these incorrect solutions and manually fix them. Specifically, for some tasks, LLMs may propose an alternative solution that differs from the ground truth solution in HumanEval, e.g., using a lambda expression to process a sequence of data instead of using a loop. To avoid overfitting the ground truth, the authors manually located and fixed errors following the problem-solving direction of the LLM instead of simply comparing the LLM-generated code with the ground truth. 

We performed multiple rounds of open coding and iterative refinement to analyze the characteristics of the located errors. Specifically, we analyzed these errors alongside two dimensions---the {\em semantic characteristics} and {\em syntactic characteristics} of these errors: 

\begin{itemize}[leftmargin=*]
    \item \textbf{Semantic characteristics} can help identify the high-level root causes of these code generation errors. Representative semantic characteristics include \textit{missing condition}, \textit{wrong (logical) direction}, \textit{incorrect condition}, etc. Analyzing these semantic characteristics can help understand the limitations of current LLMs in interpreting task requirements and generating semantically correct programs.

    \item \textbf{Syntactic characteristics} 
    can help localize where the error occurs in an incorrect code solution. Representative syntactic characteristics include \textit{incorrect code blocks}, \textit{incorrect function arguments}, etc. 
    Understanding these characteristics allows for a better assessment of current LLMs' abilities to generate different kinds of code constructs. It can also help inform the design of new techniques for localizing and repairing code generation errors made by LLMs.

\end{itemize}

Our analysis shows that the majority of code generation errors involve multiple lines of code, rather than simple errors. These errors often require substantial code restructuring and repair rather than simple fixes. Furthermore, while the overall distribution of the syntactic characteristics of these errors (i.e., error locations) is similar across different LLMs, the semantic characteristics of the errors (i.e., root causes)  vary significantly for different LLMs even for the same task. Interestingly, most of the incorrect code solutions are compilable and runnable without any compilation errors. Thus, we cannot easily capture these errors via compiler check. Careful code review and high-quality test cases are necessary to capture these errors. This also implies that modern LLMs have adequately learned the syntax rules of programming languages, but struggle with understanding intricacies in natural language task descriptions and generating delicate code with sophisticated logic. 

In summary, this paper makes the following contributions:
\begin{itemize}[leftmargin=*]
    \item We established a taxonomy of both syntactic and semantic characteristics of code generation errors through open coding and thematic analysis. Our labeling results are available at a GitHub repository~\cite{ourwebsite}.

    \item We analyzed the similarities and differences in the errors made by different LLMs, as well as the bug-fixing effort, the impact of task complexity, and the correlation between test pass rates and different kinds of errors.

    \item We discussed the implications and future opportunities for improving LLMs for code generation.

    \item We developed an interactive data analysis website to help researchers and developers examine and explore code generation errors in different categories. The website is available at \href{https://llm-code-errors.cs.purdue.edu}{https://llm-code-errors.cs.purdue.edu}.
\end{itemize}

\section{Methodology}


\subsection{Research Questions}
\label{subsec:rq}

This study investigates the following research questions.

\begin{itemize}[leftmargin=*]

    \item {\em RQ1: \rqone} This question aims to uncover the common characteristics and distinctions of code generation errors made by different LLMs. This can help us understand whether it is feasible to develop generic methods to improve LLMs or whether these models need specialized treatment.
    
    

    \item {\em RQ2: \rqfour} In practice, it is unrealistic to expect LLMs to generate fully correct code for every possible scenario. Existing studies show that some incorrect code can still serve as a useful starting point for developers~\cite{vaithilingam2022expectation, barke2023grounded}. Thus, it is important to understand what efforts are needed to fix the incorrect solutions and whether it is possible to automate the repair. This question aims to fill this knowledge gap.

    \item {\em RQ3: \rqfive} Intuitively, complex tasks are more challenging to solve than simple tasks. Yet it is unclear whether different LLMs exhibit different code generation capabilities when solving tasks of different complexity levels. Specifically, it would be useful to find out whether there is an upper bound on the complexity of tasks that LLMs can eloquently solve, which can then be used to guide or estimate the effort required for code review, testing, and repair.


    \item {\em RQ4: \rqseven} This question explores the distinctions between code that fails a subset of test cases and code that fails all test cases. It can offer insights into the specific challenges faced in achieving full correctness.
\end{itemize}

\subsection{Code Generation LLMs}

In this study, we focus on six representative code generation LLMs: CodeGen-16B~\cite{nijkamp2022codegen}, InCoder-1.3B~\cite{fried2022incoder}, GPT-3.5~\cite{chatgpt2023}, GPT-4~\cite{openai2023gpt4}, SantaCoder~\cite{allal2023santacoder}, and StarCoder~\cite{li2023starcoder}. As shown in Table~\ref{table:model_table}, these models cover a wide range of model sizes
and model performance.
\responseline{CodeGen-16B was trained on 217GB Python code from BigPython~\cite{nijkamp2022codegen}. InCoder-1.3B was trained on 159GB of open-source repositories from GitHub, GitLab, and StackOverflow. SantaCoder and StarCoder were trained on The Stack dataset~\cite{kocetkov2022stack}. The training data of GPT-3.5 and GPT-4 are currently unknown.} As GPT-3.5 and GPT-4 are constantly evolving, we used GPT-3.5-Turbo-0301 and GPT-4-0314, the two most recent model checkpoints at the time of our analysis. 

\subsection{Collection of Incorrect Code Solutions}
\label{sec:collection}

In this study, we utilize the widely used HumanEval benchmark~\cite{chen2021evaluating} to collect code generation errors made by LLMs. HumanEval includes 164 hand-written Python programming tasks, each accompanied by an average of 7.7 unit tests. These tasks involve language comprehension, reasoning, algorithms, and simple mathematics. \responseline{For each task, we followed the common practice in benchmarking the performance of code LLMs~\cite{nijkamp2022codegen, fried2022incoder, li2023starcoder} to prompt each LLM with the original prompt from HumanEval, which includes a task description and several exemplary test cases (2.7 on average). \responseline{While there are more advanced prompting strategies to augment LLMs for code generation, we are more interested in the innate capability of LLMs as the first step to understanding their limitations. Nevertheless, we discuss this as a threat to validity in Sec.~\ref{sec:threats}.} We also used greedy decoding with the temperature set to 0 to ensure the reproducibility of our results.} Then, we executed the test cases to identify incorrect solutions. We also performed a round of manual checks to find solutions that pass test cases but are not fully correct since some tasks may not have sufficient test cases. \responseline{We found 19 such cases. Example~\ref{exp-incorrect-pass-all-test} shows an incorrect solution generated by GPT-3.5, which fails to handle the case where \texttt{x} is 0. In this scenario, the output should be \texttt{"0"} instead of an empty string. However, such test cases are absent in the HumanEval benchmark.} In the end, we identified a total of 557 incorrect code solutions generated by the six models. Table~\ref{table:model_table} shows the distribution.

\begin{lstlisting}[language=Python, escapechar=!, numberstyle=\tiny\color{lightgray}, caption=\responseline{Incorrect solution that passed all test cases}, label=exp-incorrect-pass-all-test]
# [Task 44] Change numerical base of input x to base. 
def change_base(x, base):
    result = ""
    while x > 0: result, x = str(x%base) + result, x//base
    return result
\end{lstlisting}
\vspace{-5pt}

\begin{table}[t]
    \centering
    \vspace{-5pt}
    \caption{Code generation LLMs used in this study}
    \label{table:model_table}
    \scriptsize    
    \begin{tabular}{|l|r|c|c|c|}
    \hline
    \multicolumn{1}{|l|}{\multirow{2}{*}{\textbf{Model}}} & \multicolumn{1}{c|}{\multirow{2}{*}{\textbf{Release}}} & \multicolumn{1}{c|}{\multirow{2}{*}{\textbf{Size}}} & \multicolumn{2}{c|}{\textbf{Performance}}                                                         \\ \cmidrule{4-5} 
    &            &              & \multicolumn{1}{c|}{\textit{Pass@1}} & \multicolumn{1}{c|}{\textit{Incorrect Solutions}} \\ \hline\hline
    CodeGen-16B~\cite{nijkamp2022codegen}                            &         Mar. 2022                        & 16B         &            32.9\%                 &           110                   \\
    InCoder-1.3B~\cite{fried2022incoder}                         &        Apr. 2022                                     &     1.3B       &   12.2\%               &          144                \\
    GPT-3.5~\cite{chatgpt2023}                           &             Nov. 2022                & 175B                &            73.2\%                &               42                 \\
    GPT-4~\cite{openai2023gpt4}                           &             Mar. 2023                & 1.7T                &            89.0\%                &               18                 \\
    SantaCoder~\cite{allal2023santacoder}                           &             Apr. 2023          & 1.1B                      &            14.6\%                &               139                 \\
    StarCoder~\cite{li2023starcoder}                           &             May. 2023               & 15.5B                  &            34.1\%                &               104                 \\
    \hline
    \end{tabular}
    \vspace{-15pt}
\end{table}

\subsection{Manual Analysis of Incorrect Code Solutions}
We performed open coding~\cite{strauss2004open, antoine2021interaction, lazar2017research, braun2006using} to analyze the characteristics of the 557 incorrect code solutions and developed a taxonomy of code generation errors made by LLMs. 

\begin{table*}[htbp]
    \renewcommand{\arraystretch}{1}
    \centering
    \scriptsize
    \vspace{-8pt}
    \caption{Taxonomy of semantic characteristics of code generation errors made by LLMs.}
    \setlength\fboxsep{0pt}
    \begin{tabular}{|l|l|p{5.5cm}|p{5.5cm}|}
    \hline
    \multicolumn{2}{|c|}{\textbf{Error Characteristic}} & \multicolumn{1}{c|}{\textbf{Example of Incorrect Solutions}} & \multicolumn{1}{c|}{\textbf{Correct Solution}} \\ 
    \hline\hline
    \multirow{1}{*}{Condition Error} & Missing condition & 
    {
    \lstset{style=tblstyle, aboveskip=-3pt, belowskip=-8pt}
    \begin{lstlisting}[language=Python, escapechar=^, numberstyle=\tiny\color{lightgray}]
    # [Task 151] CodeGen-16b
    def double_the_difference(lst):
        sum = 0
        for i in lst:
            ^\colorbox{lightred}{\codeoperation{if} i > 0 \codeoperation{and} i \% 2 != 0:}^
                sum += i ** 2
        return sum
    \end{lstlisting}
    } &
    {
    \lstset{style=tblstyle, aboveskip=-3pt, belowskip=-8pt}
    \begin{lstlisting}[language=Python, escapechar=!, numberstyle=\tiny\color{lightgray}]
    # [Task 151] Ground Truth
    def double_the_difference(lst):
        ans = 0
        for num in lst:
            !\colorbox{lightred}{\codeoperation{if} num \% 2 == 1 \codeoperation{and} num >}! !\colorbox{lightred}{0 \codeoperation{and} \codestring{"."} \codeoperation{not} \codeoperation{in}}! !\colorbox{lightred}{\codeoperation{str}(num):}! ans += num ** 2
        return ans
    \end{lstlisting}
    }
    \\
    \cmidrule{2-4}
    ~ & Incorrect condition & 
    {
    \lstset{style=tblstyle, aboveskip=-3pt, belowskip=-8pt}
    \begin{lstlisting}[language=Python, escapechar=!, numberstyle=\tiny\color{lightgray}]
    # [Task 44] GPT-3.5
    def change_base(x, base):
        r = ""
        !\colorbox{lightred}{\codeoperation{while} x > 0:}!
            rem, r = x % base, str(rem)+r
            x = x // base
        return result
    \end{lstlisting}
    } & 
    {
    \lstset{style=tblstyle, aboveskip=-3pt, belowskip=-8pt}
    \begin{lstlisting}[language=Python, escapechar=^, numberstyle=\tiny\color{lightgray}]
    # [Task 44] Ground Truth
    def change_base(x, base):
        if x == 0: return "0"
        ret = ""
        ^\colorbox{lightred}{\codeoperation{while} x != 0:}^
            ret, x = str(x%base) + ret, x // base
        return ret
    \end{lstlisting}
    }
    \\ 
    \hline
    \multirow{1}{*}{Constant Value Error} & Constant value error &
    {
    \lstset{style=tblstyle, aboveskip=-3pt, belowskip=-8pt}
    \begin{lstlisting}[language=Python, escapechar=^, numberstyle=\tiny\color{lightgray}]
    # [Task 51] SantaCoder
    def remove_vowels(text):
        vowels = ^\colorbox{lightred}{\codestring{'aeiou'}}^
        return ''.join([char for char in text if char not in vowels])
    \end{lstlisting}
    } &
    {
    \lstset{style=tblstyle, aboveskip=-3pt, belowskip=-8pt}
    \begin{lstlisting}[language=Python, escapechar=^, numberstyle=\tiny\color{lightgray}]
    # [Task 51] Ground Truth
    def remove_vowels(text):
        return "".join(list(filter(lambda ch: ch not in ^\colorbox{lightred}{\codestring{"aeiouAEIOU"}}^, text)))
    \end{lstlisting}
    }
    \\ 
    \hline
    \multirow{1}{*}{Reference Error} & \responseline{Wrong method/variable} &
    {
    \lstset{style=tblstyle, aboveskip=-3pt, belowskip=-8pt}
    \begin{lstlisting}[language=Python, escapechar=^, numberstyle=\tiny\color{lightgray}]
    ^\responsecodecommentref{}\# [Task 79] InCoder-1B^
    ^\responsecodeoperationref{}def^ ^\responsecodemainref{}decimal\_to\_binary(decimal):^
        ^\responsecodeoperationref{}return^ ^\colorbox{lightred}{{\responsecodeoperationref{}{str}}{\responsecodemainref{}(decimal)}}^ ^\responsecodemainref{}+^ ^\responsecodestringref{}"db"^
    \end{lstlisting}
    } &
    {
    \lstset{style=tblstyle, aboveskip=-3pt, belowskip=-8pt}
    \begin{lstlisting}[language=Python, escapechar=^, numberstyle=\tiny\color{lightgray}]
    ^\responsecodecommentref{}\# [Task 79] Ground Truth^
    ^\responsecodeoperationref{}def^ ^\responsecodemainref{}decimal\_to\_binary(decimal):^
        ^\responsecodeoperationref{}return^ ^\colorbox{lightred}{{\responsecodeoperationref{}{bin}}{\responsecodemainref{}(decimal)}}^ ^\responsecodemainref{}+^ ^\responsecodestringref{}"db"^
    \end{lstlisting}
    }
    \\ 
    \cmidrule{2-4}
    ~ & Undefined name & {
    \lstset{style=tblstyle, aboveskip=-3pt, belowskip=-8pt}
    \begin{lstlisting}[language=Python, escapechar=^, numberstyle=\tiny\color{lightgray}]
    # [Task 83] CodeGen-16B
    def starts_one_ends(n):
        if n == 1: return 1
        return starts_one_ends(n-1)+^\colorbox{lightred}{ends\_one\_starts(n-1)}^
    \end{lstlisting}
    } &
    {
    \lstset{style=tblstyle, aboveskip=-3pt, belowskip=-8pt}
    \begin{lstlisting}[language=Python, escapechar=^, numberstyle=\tiny\color{lightgray}]
    # [Task 83] Ground Truth
    def starts_one_ends(n):
        if n == 1: return 1
        return 18 * 10 ** (n - 2)
    \end{lstlisting}
    } \\ 
    \hline
    \multirow{2}{*}{\begin{tabular}{@{}l@{}}
         Operation/ \\
         Calculation Error
    \end{tabular}} & Incorrect arithmetic operation &
    {
    \lstset{style=tblstyle, aboveskip=-3pt, belowskip=-8pt}
    \begin{lstlisting}[language=Python, escapechar=^, numberstyle=\tiny\color{lightgray}]
    # [Task 44] InCoder-1B
    def change_base(x: int, base: int):
        digits = []
        while x:
            digits.append(str(x % base))
            ^\colorbox{lightred}{x /= base}^
        return ''.join(reversed(digits))
    \end{lstlisting}
    } &
    {
    \lstset{style=tblstyle, aboveskip=-3pt, belowskip=-8pt}
    \begin{lstlisting}[language=Python, escapechar=^, numberstyle=\tiny\color{lightgray}]
    # [Task 44] Ground Truth
    def change_base(x: int, base: int):
        if x == 0: return "0"
        ret = ""
        while x != 0:
            ret = str(x % base) + ret
            ^\colorbox{lightred}{x //= base}^
        return ret
    \end{lstlisting}
    }\\ 
    \cmidrule{2-4}
     & \responseline{Incorrect comparison operation} & {
    \lstset{style=tblstyle, aboveskip=-3pt, belowskip=-8pt}
    \begin{lstlisting}[language=Python, escapechar=^, numberstyle=\tiny\color{lightgray}]
    ^\responsecodecommentref{}\# [Task 138] CodeGen-16B^
    ^\responsecodeoperationref{}def^ ^\responsecodemainref{}is\_equal\_to\_sum\_even(n):^
        ^\responsecodeoperationref{}return^ ^\colorbox{lightred}{\responsecodemainref{}n <= 8}^ ^\responsecodeoperationref{}and^ ^\responsecodemainref{}n \% 2 == 0^
    \end{lstlisting}
    } &
    {
    \lstset{style=tblstyle, aboveskip=-3pt, belowskip=-8pt, escapechar=^}
    \begin{lstlisting}[language=Python, numberstyle=\tiny\color{lightgray}]
    ^\responsecodecommentref{}\# [Task 138] Ground Truth^
    ^\responsecodeoperationref{}def^ ^\responsecodemainref{}is\_equal\_to\_sum\_even(n):^
        ^\responsecodeoperationref{}return^ ^\colorbox{lightred}{\responsecodemainref{}n >= 8}^ ^\responsecodeoperationref{}and^ ^\responsecodemainref{}n \% 2 == 0^
    \end{lstlisting}
    }\\ 
    \hline
    \multirow{1}{*}{Garbage Code} & Only comments &
    {
    \lstset{style=tblstyle, aboveskip=-3pt, belowskip=-8pt}
    \begin{lstlisting}[language=Python, escapechar=^, numberstyle=\tiny\color{lightgray}]
    # [Task 152] InCoder-1B
    def compare(game,guess):
        ^\colorbox{lightred}{\codecomment{\# Regenerate the task description.}}^
    \end{lstlisting}
    } &
    {
    \lstset{style=tblstyle, aboveskip=-3pt, belowskip=-8pt}
    \begin{lstlisting}[language=Python, escapechar=^, numberstyle=\tiny\color{lightgray}]
    # [Task 152] Ground Truth
    def compare(game,guess):
        return [abs(game[i] - guess[i]) for i in range(len(game))]
    \end{lstlisting}
    } \\ 
    \cmidrule{2-4}
    ~ & Meaningless code snippet &
    {
    \lstset{style=tblstyle, aboveskip=-3pt, belowskip=-8pt}
    \begin{lstlisting}[language=Python, escapechar=^, numberstyle=\tiny\color{lightgray}]
    # [Task 138] StarCoder
    def is_equal_to_sum_even(n):
        ^\colorbox{lightred}{\codeoperation{pass}}^
    \end{lstlisting}
    } &
    {
    \lstset{style=tblstyle, aboveskip=-3pt, belowskip=-8pt}
    \begin{lstlisting}[language=Python, escapechar=^, numberstyle=\tiny\color{lightgray}]
    # [Task 138] Ground Truth
    def is_equal_to_sum_even(n):
        return n >= 8 and n % 2 == 0
    \end{lstlisting}
    } \\ 
    \cmidrule{2-4}
    ~ & Wrong (logical) direction &
    {
    \lstset{style=tblstyle, aboveskip=-3pt, belowskip=-8pt}
    \begin{lstlisting}[language=Python, escapechar=^, numberstyle=\tiny\color{lightgray}]
    # [Task 20] InCoder-1.3B
    def find_closest_elements(numbers):
        closest_to_one, closest_to_two = numbers[0], numbers[1]
        for number in numbers:
            ^\colorbox{lightred}{\codeoperation{if} number < closest\_to\_one:}^
                ^\colorbox{lightred}{closest\_to\_one = number}^
            ^\colorbox{lightred}{\codeoperation{if} number > closest\_to\_two:}^
                ^\colorbox{lightred}{cloeset\_to\_two = number}^
        return closest_to_one, closest_to_two
    \end{lstlisting}
    } &
    {
    \lstset{style=tblstyle, aboveskip=-3pt, belowskip=-8pt}
    \begin{lstlisting}[language=Python, escapechar=^, numberstyle=\tiny\color{lightgray}]
    # [Task 20] InCoder-1.3B
    def find_closest_elements(numbers):
        min_diff, min_pair = float("inf"), None
        for l, r in zip(numbers[:-1], numbers[1:]):
            ^\colorbox{lightred}{diff = r - l}^
            ^\colorbox{lightred}{\codeoperation{if} diff < min\_diff:}^
                ^\colorbox{lightred}{min\_diff = diff}^
                ^\colorbox{lightred}{min\_pair = (l, r)}^
        return min_pair
    \end{lstlisting}
    } \\ 
    \hline
    \multirow{2}{*}{\begin{tabular}{@{}l@{}}
         Incomplete Code/ \\
         Missing Steps
    \end{tabular}} & {Missing one step} \vspace{-8pt} &
    {
    \lstset{style=tblstyle, aboveskip=-3pt, belowskip=-8pt}
    \begin{lstlisting}[language=Python, escapechar=^, numberstyle=\tiny\color{lightgray}]
    # [Task 16] InCoder-1B
    def count_distinct_chars(string):
        return len(set(string))
    \end{lstlisting}
    } &
    {
    \lstset{style=tblstyle, aboveskip=-3pt, belowskip=-8pt}
    \begin{lstlisting}[language=Python, escapechar=^, numberstyle=\tiny\color{lightgray}]
    # [Task 16] Ground Truth
    def count_distinct_chars(string):
        return len(set(^\colorbox{lightred}{string.lower()}^))
    \end{lstlisting}
    } \\ 
    \cmidrule{2-2}
     & Missing multiple steps &  & \\ 
    \hline
    \multirow{1}{*}{Memory Error} & Infinite loop &
    {
    \lstset{style=tblstyle, aboveskip=-3pt, belowskip=-8pt}
    \begin{lstlisting}[language=Python, escapechar=^, numberstyle=\tiny\color{lightgray}]
    # [Task 100] CodeGen-16b
    def make_a_pile(n):
        if n % 2 == 0: 
            return [n] + ^\colorbox{lightred}{make\_a\_pile(n+2)}^
        else: 
            return [n] + ^\colorbox{lightred}{make\_a\_pile(n+1)}^
    \end{lstlisting}
    } &
    {
    \lstset{style=tblstyle, aboveskip=-3pt, belowskip=-8pt}
    \begin{lstlisting}[language=Python, escapechar=^, numberstyle=\tiny\color{lightgray}]
    # [Task 100] Ground Truth
    def make_a_pile(n):
        ans, num = [], n
        for _ in range(n):
            ans.append(num)
            num += 2
        return ans
    \end{lstlisting}
    }\\ 
    \hline
    \end{tabular}
    \vspace{-17pt}
    \label{tab:semantic_taxonomy}
\end{table*}

\vspace{0.25mm}
\smalltitle{Open coding.} From the 557 incorrect solutions, we first randomly sampled 160 of them as a starting point for analysis. The sample size is statistically significant, with a 90\% confidence level and a \responseline{5.5\%} margin of error. Two authors independently identified the erroneous parts of each incorrect solution and documented the root causes of the errors. \responseline{For incorrect code solutions with multiple errors, the authors labeled the characteristics of each individual error.} Since LLMs may generate alternative solutions compared with the ground-truth solution from HumanEval, we chose to manually debug the incorrect solution rather than simply comparing it with ground truth. Specifically, the two authors executed the failed test cases and performed step-by-step debugging to locate the errors and identify their root causes. 

The authors documented all error locations and root causes and discussed them with other authors after the initial coding. They refined code labels and came up with an initial codebook. At this stage, we found that code generation errors made by LLMs can be categorized along two dimensions based on their \textit{semantic} and \textit{syntactic} characteristics. \responseline{Semantic characteristics help identify the high-level root causes of code generation errors, such as a wrong logical direction to solve the task. In contrast, syntactic characteristics assist in error localization, such as determining whether the error is in the method name or the arguments.} The initial codebook includes seven semantic characteristics and eight syntactic characteristics.

\begin{table*}[htbp]
    \renewcommand{\arraystretch}{1}
    \centering
    \scriptsize
    \setlength\fboxsep{0pt}
    \vspace{-8pt}
    \caption{Taxonomy of syntactic characteristics of code generation errors made by LLMs.}
    \begin{tabular}{|l|l|p{5.6cm}|p{5.6cm}|}
    \hline
    \multicolumn{2}{|c|}{\textbf{Error Characteristic}} & \multicolumn{1}{c|}{\textbf{Example of Incorrect Solutions}} & \multicolumn{1}{c|}{\textbf{Correct Solution}} \\ 
    \hline\hline
    \multirow{1}{*}{Conditional Error} & If error & 
    {
    \lstset{style=tblstyle, aboveskip=-3pt, belowskip=-8pt}
    \begin{lstlisting}[language=Python, escapechar=^, numberstyle=\tiny\color{lightgray}]
    # [Task 151] CodeGen-16b
    def double_the_difference(lst):
        sum = 0
        for i in lst:
            ^\colorbox{lightred}{\codeoperation{if} i > 0 \codeoperation{and} i \% 2 != 0:}^ 
                sum += i ** 2
        return sum
    \end{lstlisting}
    } &
    {
    \lstset{style=tblstyle, aboveskip=-3pt, belowskip=-8pt}
    \begin{lstlisting}[language=Python, escapechar=!, numberstyle=\tiny\color{lightgray}]
    # [Task 151] Ground Truth
    def double_the_difference(lst):
        ans = 0
        for num in lst:
            !\colorbox{lightred}{\codeoperation{if} num\%2==1 \codeoperation{and} num>0 \codeoperation{and} \codestring{"."} \codeoperation{not}}! 
                  !\colorbox{lightred}{\codeoperation{in str}(num): ans += num ** 2}!
        return ans
    \end{lstlisting}
    } \\ 
    \hline
    \multirow{1}{*}{Loop Error} & For error \vspace{-8pt} & 
    {
    \lstset{style=tblstyle, aboveskip=-3pt, belowskip=-8pt}
    \begin{lstlisting}[language=Python, escapechar=^, numberstyle=\tiny\color{lightgray}]
    # [Task 121] GPT-3.5
    def solution(lst):
        sum = 0
        ^\colorbox{lightred}{\codeoperation{for} i \codeoperation{in} \codeoperation{range}(1, \codeoperation{len}(lst), 2):}^
            if lst[i] % 2 != 0: sum +=lst[i]
        return sum
    \end{lstlisting}
    } &
    {
    \lstset{style=tblstyle, aboveskip=-3pt, belowskip=-8pt}
    \begin{lstlisting}[language=Python, escapechar=!, numberstyle=\tiny\color{lightgray}]
    # [Task 121] Ground Truth
    def solution(lst):
        return sum([x !\colorbox{lightred}{\codeoperation{for} idx, x \codeoperation{in} }!
               !\colorbox{lightred}{\codeoperation{enumerate}(lst)}! if idx%2==0 and x%2==1])
    \end{lstlisting}
    } \\ 
    \cmidrule{2-2}
    ~ & While error & 
    {
    } &
    {
    } \\ 
    \hline
    \multirow{1}{*}{Return Error} & Incorrect return value & 
    {
    \lstset{style=tblstyle, aboveskip=-3pt, belowskip=-8pt}
    \begin{lstlisting}[language=Python, escapechar=^, numberstyle=\tiny\color{lightgray}]
    # [Task 103] GPT-3.5
    def rounded_avg(n, m)
       if n > m: return -1
       avg=round(sum(range(n,m+1))/(m-n+1))
       return ^\colorbox{lightred}{\codeoperation{bin}(avg)[2:]}^
    \end{lstlisting}
    } &
    {
    \lstset{style=tblstyle, aboveskip=-3pt, belowskip=-8pt}
    \begin{lstlisting}[language=Python, escapechar=^, numberstyle=\tiny\color{lightgray}]
    # [Task 103] Ground Truth
    def rounded_avg(n, m)
        if n > m: return -1
        avg = round((n + m) / 2)
        return ^\colorbox{lightred}{\codeoperation{bin}(avg)}^
    \end{lstlisting}
    } \\ 
    \hline
    \multirow{1}{*}{Method Call Error} & \responseline{Incorrect function name} \vspace{-11pt} & 
    {
    \lstset{style=tblstyle, aboveskip=-3pt, belowskip=-8pt}
    \begin{lstlisting}[language=Python, escapechar=^, numberstyle=\tiny\color{lightgray}]
    ^\responsecodecommentref{}\# [Task 54] StarCoder^
    ^\responsecodeoperationref{}def^ ^\responsecodemainref{}same\_chars(s0, s1):^
        ^\responsecodeoperationref{}return^ ^\colorbox{lightred}{{\responsecodeoperationref{}{sorted}}\responsecodemainref{}(s0)}^ ^\responsecodemainref{}==^ ^\colorbox{lightred}{{\responsecodeoperationref{}{sorted}}\responsecodemainref{}(s1)}^
    \end{lstlisting}
    } &
    {
    \lstset{style=tblstyle, aboveskip=-3pt, belowskip=-8pt}
    \begin{lstlisting}[language=Python, escapechar=^, numberstyle=\tiny\color{lightgray}]
    ^\responsecodecommentref{}\# [Task 54] Ground Truth^
    ^\responsecodeoperationref{}def^ ^\responsecodemainref{}same\_chars(s0, s1):^
        ^\responsecodeoperationref{}return^ ^\colorbox{lightred}{{\responsecodeoperationref{}{set}}\responsecodemainref{}(s0)}^ ^\responsecodemainref{}==^ ^\colorbox{lightred}{{\responsecodeoperationref{}{set}}\responsecodemainref{}(s1)}^
    \end{lstlisting}
    } \\ 
    \cmidrule{2-2}
    ~ & Incorrect function arguments & & \\ 
    \cmidrule{2-2}
    ~ & Incorrect method call target & &\\ 
    \hline
    \multirow{1}{*}{Assignment Error} & {Incorrect constant} \vspace{-11pt} & 
    {
    \lstset{style=tblstyle, aboveskip=-3pt, belowskip=-8pt}
    \begin{lstlisting}[language=Python, escapechar=^, numberstyle=\tiny\color{lightgray}]
    # [Task 138] InCoder-1.3B
    def is_equal_to_sum_even(n):
        return n >=^\colorbox{lightred}{ 4 }^ and n % 2 == 0
    \end{lstlisting}
    } &
    {
    \lstset{style=tblstyle, aboveskip=-3pt, belowskip=-8pt}
    \begin{lstlisting}[language=Python, escapechar=^, numberstyle=\tiny\color{lightgray}]
    # [Task 138] Ground Truth
    def is_equal_to_sum_even(n):
        return n >=^\colorbox{lightred}{ 8 }^ and n % 2 == 0
    \end{lstlisting}
    } \\ 
    \cmidrule{2-2}
    ~ & Incorrect arithmetic & & \\ 
    \cmidrule{2-2}
    ~ & Incorrect variable name & & \\ 
    \cmidrule{2-2}
    ~ & Incorrect comparison & & \\ 
    \hline
    Import Error & Import error & 
    {
    \lstset{style=tblstyle, aboveskip=-3pt, belowskip=-8pt}
    \begin{lstlisting}[language=Python, escapechar=^, numberstyle=\tiny\color{lightgray}]
    # [Task 133] StarCoder
    def sum_squares(lst):
        return sum[int(^\colorbox{lightred}{math.ceil}^(i)**2 for i in lst]
    \end{lstlisting}
    } &
    {
    \lstset{style=tblstyle, aboveskip=-3pt, belowskip=-8pt}
    \begin{lstlisting}[language=Python, escapechar=^, numberstyle=\tiny\color{lightgray}]
    # [Task 133] Ground Truth
    def sum_squares(lst):
        ^\colorbox{lightred}{\codeoperation{import} math}^
        return sum(map(lambda x: math.ceil(x)**2, lst))
    \end{lstlisting}
    } \\ 
    \hline
    \multirow{1}{*}{Code Block Error} & Incorrect code block &
    {
    \lstset{style=tblstyle, aboveskip=-3pt, belowskip=-8pt}
    \begin{lstlisting}[language=Python, escapechar=^, numberstyle=\tiny\color{lightgray}]
    # [Task 83] InCoder-1.3B
    def starts_one_ends(n):
        ^\colorbox{lightred}{count = 0}^
        ^\colorbox{lightred}{\codeoperation{while} n > 0:}^
            ^\colorbox{lightred}{count, n = count + 1, n / 10}^
        ^\colorbox{lightred}{\codeoperation{return} count}^

    \end{lstlisting}
    } &
    {
    \lstset{style=tblstyle, aboveskip=-3pt, belowskip=-8pt}
    \begin{lstlisting}[language=Python, escapechar=^, numberstyle=\tiny\color{lightgray}]
    # [Task 83] Ground Truth
    def starts_one_ends(n):
        if n == 1: 
            return 1
        return 18 * 10 ** (n - 2)
    \end{lstlisting}
    } \\ 
    \cmidrule{2-4}
    ~ & Missing code block &
    {
    \lstset{style=tblstyle, aboveskip=-3pt, belowskip=-8pt}
    \begin{lstlisting}[language=Python, escapechar=^, numberstyle=\tiny\color{lightgray}]
    # [Task 60] CodeGen-16B
    def next_smallest(lst):
        if len(lst)<2: return None
        lst.sort()
        return lst[1]
    \end{lstlisting}
    } &
    {
    \lstset{style=tblstyle, aboveskip=-3pt, belowskip=-8pt}
    \begin{lstlisting}[language=Python, escapechar=^, numberstyle=\tiny\color{lightgray}]
    # [Task 60] Ground Truth
    def is_prime(n):
        if len(lst)<=1: return None
        sorted_list=sorted(lst)
        ^\colorbox{lightred}{\codeoperation{for} x \codeoperation{in} sorted\_list:}^
            ^\colorbox{lightred}{\codeoperation{if} x!=sorted\_list[0]: \codeoperation{return} x}^
    \end{lstlisting}
    } \\ 
    \hline
    \end{tabular}
    \vspace{-17pt}
    \label{tab:syntactic_taxonomy}
\end{table*}

\vspace{0.25mm}
\smalltitle{Iterative refinement of the codebook.} After obtaining the initial codebook, we invited another two authors to iteratively improve the codebook. 
The four authors first independently analyzed 10 incorrect code snippets following the same procedure described above and labeled the error characteristics based on the initial codebook. 
If a new characteristic was identified, an author created a new label to describe the characteristic.

After the first round of labeling, we computed Fleiss’ Kappa~\cite{fleiss1971measuring} to measure the inter-rater agreement. We used Fleiss' Kappa instead of Cohen's Kappa, since we had more than two labelers and more than two labels. The initial scores were 0.37 and 0.32 for semantic characteristics and syntactic characteristics, respectively~\cite{landis1977measurement}. To figure out where the disagreements were, the four authors met to discuss the disagreements and exchanged opinions about updating the codebook. They found that the low agreement was due to missing error characteristics in the initial codebook.

The four authors then refined the codebook with \responseline{11 semantic characteristics and 13 syntactic characteristics} and labeled another batch of 10 incorrect solutions. The Fleiss’ Kappa scores of this round of labeling were 0.68 and 0.69 for semantic characteristics and syntactic characteristics, respectively, indicating substantial agreement~\cite{landis1977measurement}. The authors further discussed the disagreements and refined the codebook with \responseline{13 semantic characteristics and 14 syntactic characteristics}. Then, they conducted the third round of labeling with a new batch of 10 errors. The authors did not find any new error characteristics in this round, and the Fleiss’ Kappa scores increased to 0.84 and 0.71. At this point, the authors believed that the codebook was comprehensive enough. The final codebook includes 13 semantic characteristics and 14 syntactic characteristics. 

\vspace{0.25mm}
\smalltitle{Analyzing the remaining dataset.} The two authors used the final codebook to label the remaining incorrect solutions. The final Fleiss' Kappa scores were 0.91 and 0.91 for semantic and syntactic characteristics, indicating perfect agreement~\cite{landis1977measurement}. \responseline{They had disagreements on 29 errors' semantic characteristics and 28 errors' syntactic characteristics. These disagreements were resolved after discussing them with all the authors.} No new error characteristics were found.
The final coding results were documented in a spreadsheet and shared on GitHub~\cite{ourwebsite}.
{The whole labeling process took about 328 person-hours.}

\subsection{Analysis of Repair Effort}

{
\responseref{}
To investigate the repair effort (RQ2), we employ three different metrics to measure the similarity between incorrect model-generated code and the corresponding correct solution. To ensure a fair comparison, we first removed all LLM-generated comments before calculation. We used Levenshtein distance~\cite{levenshtein1966binary} to compute the minimum number of edits (i.e., insertions, deletions, or substitutions) required to change an incorrect solution to the correct solution. We also used Jaccard similarity~\cite{jaccard1901etude} as another textual similarity metric. Both of them are widely used for fault localization~\cite{abreu2007accuracy, wen2019historical}.
We further used CodeBERTScore~\cite{zhou2023codebertscore} to measure the semantic similarity between the incorrectly generated code and the ground truth.

Note that for some tasks, an LLM may propose an alternative solution with errors compared with the ground-truth solution in HumanEval. We identified 17 incorrect solutions where the LLM proposed an alternative way to solve the task but did not correctly solve it. In such cases, it is unfair to directly compare the incorrect code with the ground truth. To address this issue, one author manually solved the task following the LLM's solution and computed the metrics by comparing the incorrect solution with the alternative, correct solution.
}

\begin{figure*}[t]
     \centering
     \vspace{-8pt}
     \begin{subfigure}[b]{0.26\textwidth}
         \centering
         \includegraphics[width=1\linewidth]{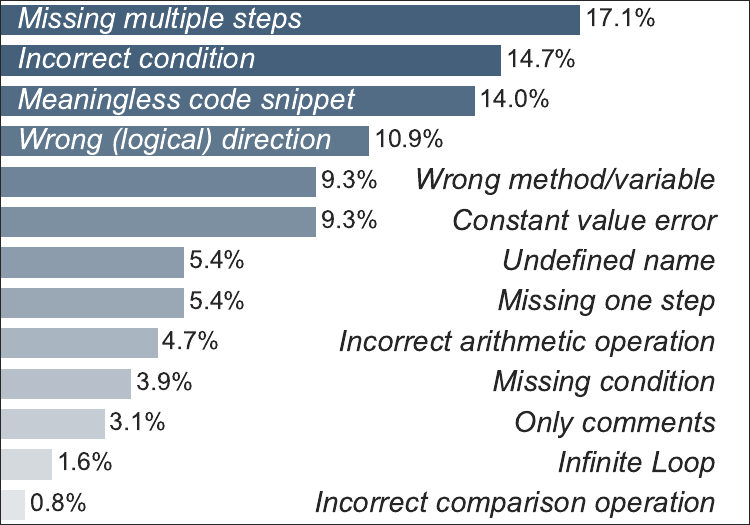}
         \vspace{-15pt}
        \caption{CodeGen-16B}
        \vspace{-5pt}
         \label{fig:codegen_semantic}
     \end{subfigure}%
     \hspace{11mm}
     \begin{subfigure}[b]{0.26\textwidth}
         \centering
         \includegraphics[width=1\linewidth]{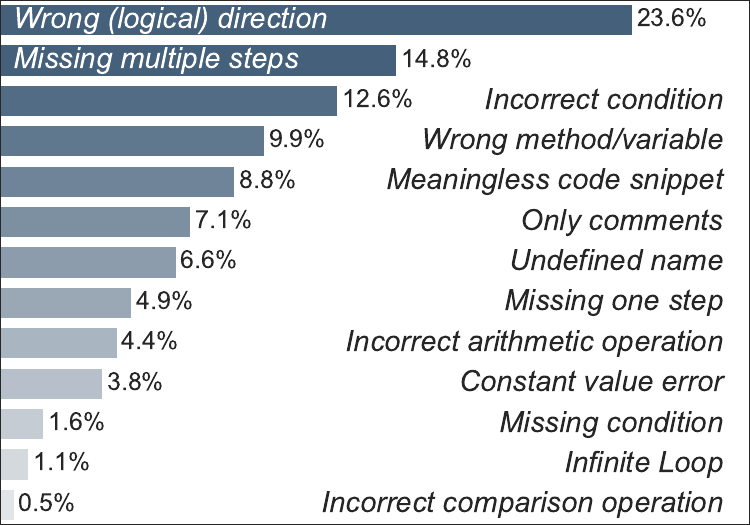}
         \vspace{-15pt}
        \caption{InCoder-1.3B}
        \vspace{-5pt}
         \label{fig:incoder_semantic}
     \end{subfigure}%
     \hspace{11mm}
     \begin{subfigure}[b]{0.26\textwidth}
         \centering
         \includegraphics[width=1\linewidth]{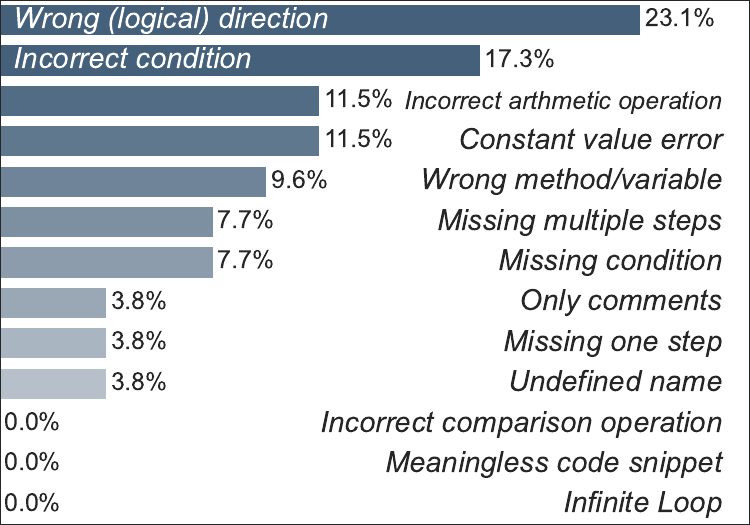}
         \vspace{-15pt}
         \caption{GPT-3.5}
         \vspace{-5pt}
         \label{fig:chatgpt_semantic}
     \end{subfigure}%
     \vfill
     \begin{subfigure}[b]{0.26\textwidth}
         \centering
         \vspace{8pt}
         \includegraphics[width=1\linewidth]{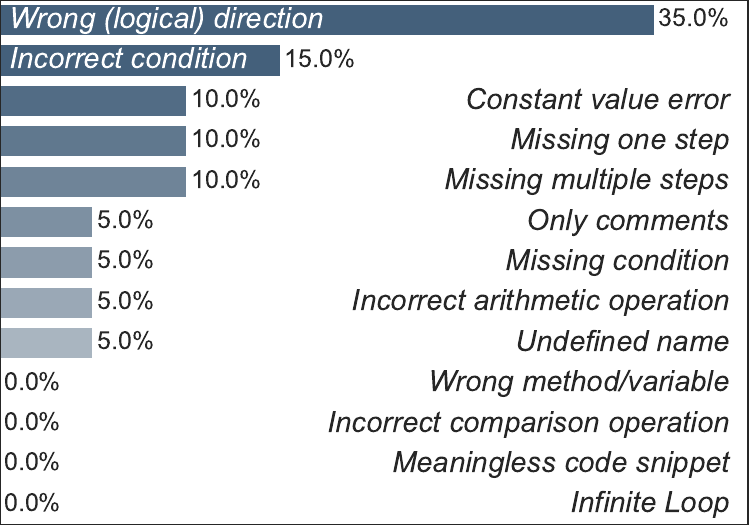}
         \vspace{-15pt}
         \caption{GPT-4}
         \label{fig:gpt4_semantic}
     \end{subfigure}%
     \hspace{11mm}
     \begin{subfigure}[b]{0.26\textwidth}
         \centering
         \vspace{8pt}
         \includegraphics[width=1\linewidth]{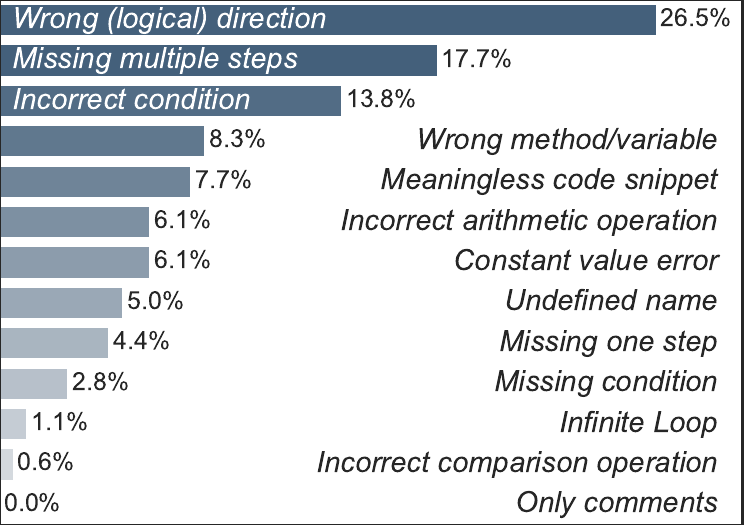}
         \vspace{-15pt}
         \caption{SantaCoder}
         \label{fig:santacoder_semantic}
     \end{subfigure}%
     \hspace{11mm}
     \begin{subfigure}[b]{0.26\textwidth}
         \centering
         \vspace{8pt}
         \includegraphics[width=1\linewidth]{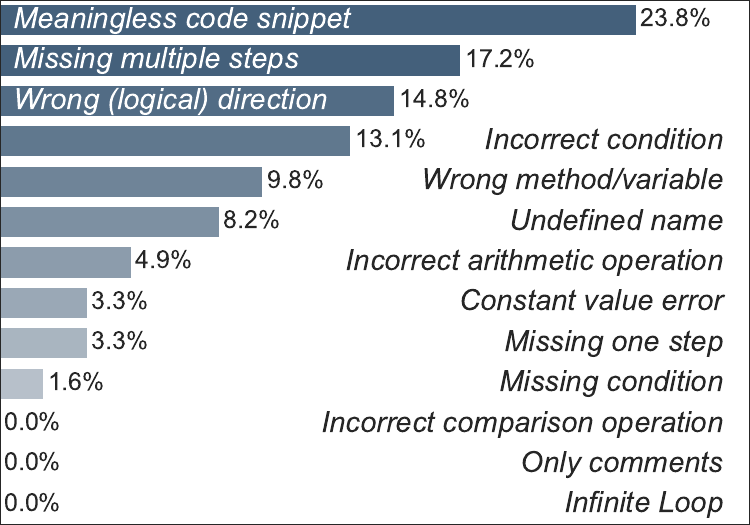}
         \vspace{-15pt}
         \caption{StarCoder}
         \label{fig:starcoder_semantic}
     \end{subfigure}%
     \vspace{-3pt}
    
    \caption{Distribution of \textit{semantic characteristics} of code generation errors made by six LLMs.}
     \vspace{-20pt}
    \label{fig:semantic}
\end{figure*}

\section{Results}
In this section, we denote the 164 programming tasks in HumanEval~\cite{chen2021evaluating} as {\ttfamily Task 0-163}. Due to the page limit, some code examples are simplified. We refer the readers to our Github repository for more details~\cite{ourwebsite}.

\subsection{RQ1: Characteristics of Code Generation Errors}
\label{subsec:rq1}

Table~\ref{tab:semantic_taxonomy} and Table~\ref{tab:syntactic_taxonomy} present the finalized taxonomy of code generation errors made by LLMs. The taxonomy categorizes code generation errors based on their semantic characteristics (i.e., root causes) and syntactic characteristics (i.e., error locations). In total, there are 13 semantic characteristics in 7 categories and 14 syntactic characteristics in 7 categories. We elaborate on each of them below.

\subsubsection{Semantic Characteristics} 

\begin{itemize}[leftmargin=*]
    \item \textbf{Condition Error} includes \textit{missing condition} and \textit{incorrect condition}. Missing condition is when a necessary condition is omitted, while incorrect condition is when an condition is incorrectly formulated in an if statements or a loop, leading to errors. 

    \item \textbf{Constant Value Error} is an error that occurs when an incorrect constant value is set, which can occur in function arguments, assignments, or other parts of the code. 
    
    \item \textbf{Reference Error} involves incorrect references to variables or functions, which includes the usage of an incorrect function or variable that does not match the requirement (\textit{wrong method/variable}) and reference to a variable or method name that has not been defined (\textit{undefined name}). 
    
    \item \textbf{Operation/Calculation Error} indicates the mistakes in mathematical or logical operations, \responseline{e.g., an incorrect comparison operation in a return statement ``\texttt{return n <= 8}.''}

    \item \textbf{Garbage Code} is defined as unnecessary or irrelevant code that does not contribute to the intended functionality. It can occur in several forms: a \textit{meaningless code snippet}, where the code, though syntactically correct, is irrelevant to the assigned task; \textit{only comments}, where the code consists exclusively of comments without any executable statements; or \textit{wrong (logical) direction}, where the code significantly deviates from the intended task logic or expected outcomes.

    \item \textbf{Incomplete Code/Missing Steps} indicates the absence of crucial steps to achieve the task.
    
    \item \textbf{Memory Error} includes \textit{infinite loop}, which is a loop or recursion that never terminates.
\end{itemize}

\smalltitle{Comparison between LLMs.}
Fig.~\ref{fig:semantic} shows the distribution of 
the 13 semantic characteristics for each LLM. We find that several characteristics are frequently shared among all LLMs, such as {\em incorrect condition} and {\em wrong (logical) direction}. This implies that all LLMs struggle with certain kinds of task requirements, such as handling complex logic conditions, regardless of their model size and capability.

However, small models such as InCoder and CodeGen are more likely to generate {\em meaningless code} and code that {\em miss multiple steps}, while larger models such as GPT-3.5 and GPT-4 tend to make more {\em constant value errors} and {\em arithmetic operation errors}. Notably, incorrect code generated by GPT-4 only exhibited 9 of the 13 semantic characteristics, while incorrect code generated by smaller models exhibited all sorts of errors. One plausible reason is that GPT-3.5 and GPT-4 are much larger and are thus better at interpreting task descriptions. 
\responseline{For instance, neither GPT-3.5 nor GPT-4 generated any meaningless code snippets. In contrast, 7\% to 25\% of the incorrect code solutions produced by the other four LLMs consist of meaningless code snippets.}

\vspace{-5pt}
\begin{finding}
    \label{finding:1}
   The most common semantic characteristics among six LLMs are \textit{wrong (logical) direction} and \textit{incorrect condition}, indicating that all LLMs struggle with interpreting complex task requirements and generating correct logic conditions. Compared with ultra-large models such as GPT-3.5, small models generate more meaningless code and code that misses multiple steps.
\end{finding}
\vspace{-5pt}

\begin{figure*}[h]
     \centering
     \vspace{-8pt}
     \begin{subfigure}[b]{0.26\textwidth}
         \centering
         \includegraphics[width=1\linewidth]{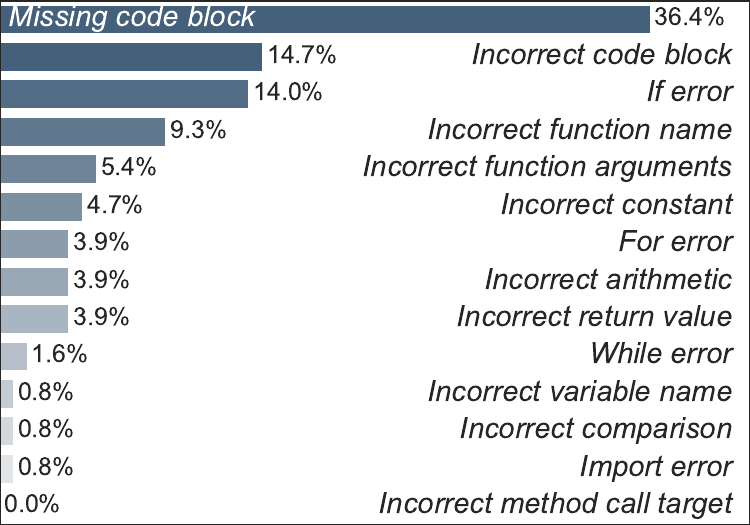}
         \vspace{-15pt}
        \caption{CodeGen-16B}
        \vspace{-5pt}
         \label{fig:codegen_syntactic}
     \end{subfigure}%
     \hspace{11mm}
     \begin{subfigure}[b]{0.26\textwidth}
         \centering
         \includegraphics[width=1\linewidth]{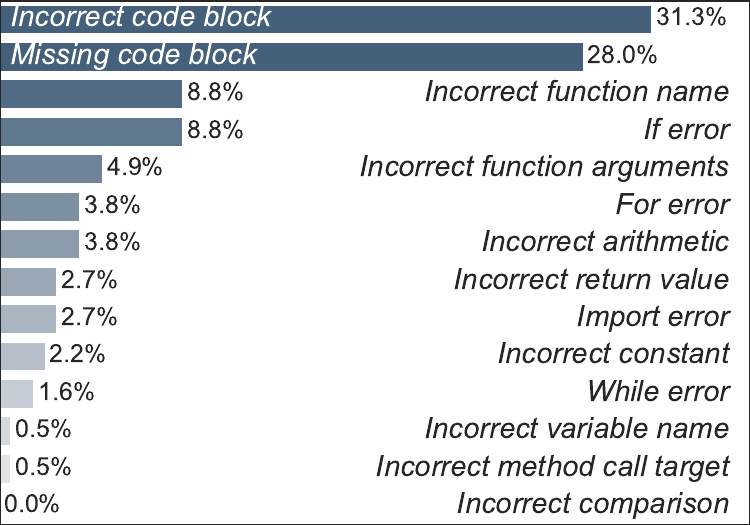}
         \vspace{-15pt}
        \caption{InCoder-1.3B}
        \vspace{-5pt}
         \label{fig:incoder_syntactic}
     \end{subfigure}%
     \hspace{11mm}
     \begin{subfigure}[b]{0.26\textwidth}
         \centering
         \includegraphics[width=1\linewidth]{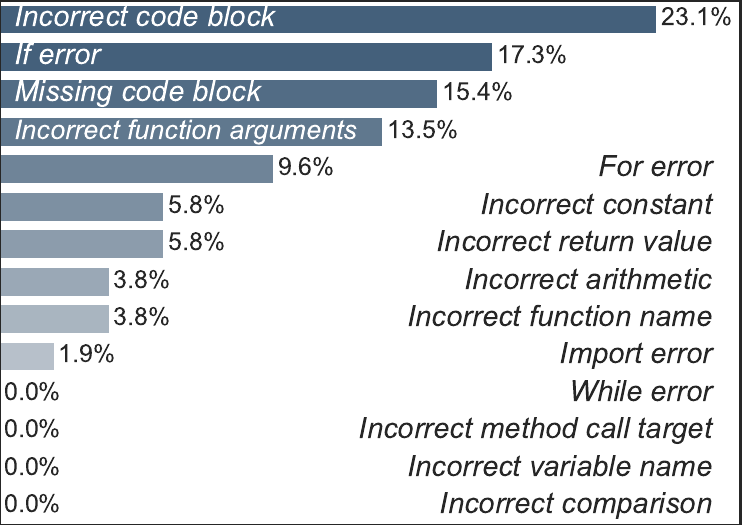}
         \vspace{-15pt}
         \caption{GPT-3.5}
         \vspace{-5pt}
         \label{fig:chatgpt_syntactic}
     \end{subfigure}%
     \vfill
     \begin{subfigure}[b]{0.26\textwidth}
         \centering
         \vspace{8pt}
         \includegraphics[width=1\linewidth]{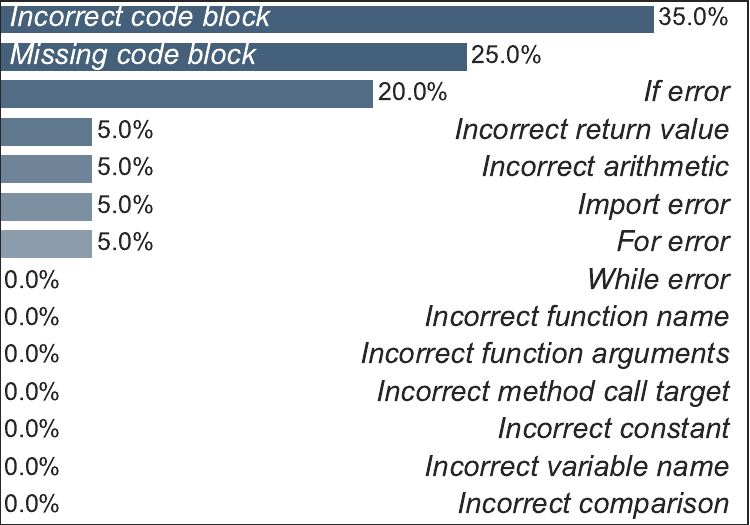}
         \vspace{-15pt}
         \caption{GPT-4}
         \label{fig:gpt4_syntactic}
     \end{subfigure}%
     \hspace{11mm}
     \begin{subfigure}[b]{0.26\textwidth}
         \centering
         \vspace{8pt}
         \includegraphics[width=1\linewidth]{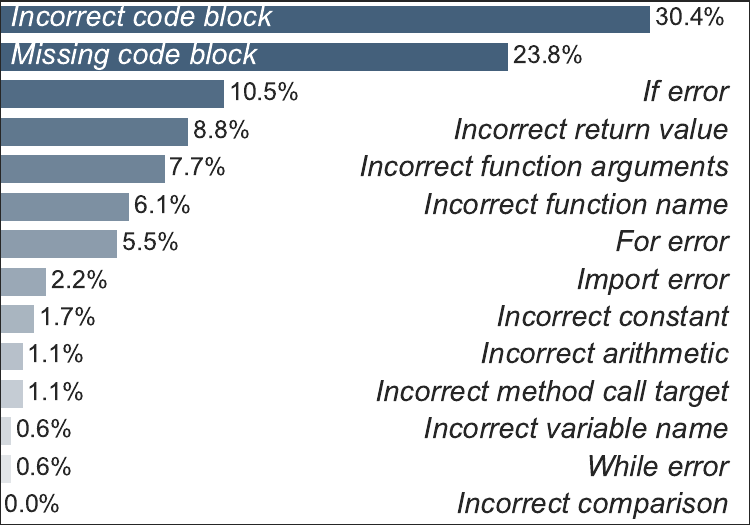}
         \vspace{-15pt}
         \caption{SantaCoder}
         \label{fig:santacoder_syntactic}
     \end{subfigure}%
     \hspace{11mm}
     \begin{subfigure}[b]{0.26\textwidth}
         \centering
         \vspace{8pt}
         \includegraphics[width=1\linewidth]{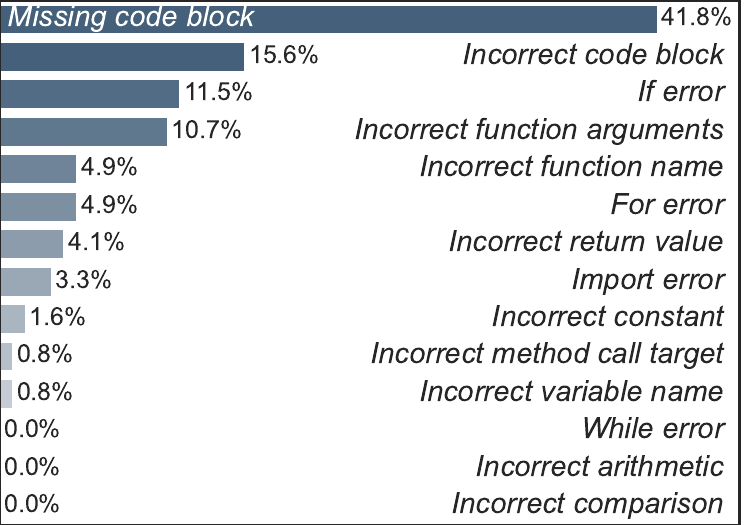}
         \vspace{-15pt}
         \caption{StarCoder}
         \label{fig:starcoder_syntactic}
     \end{subfigure}%
     \vspace{-3pt}
    \caption{Distribution of \textit{syntactic characteristics} of code generation errors made by six LLMs.}
    \vspace{-20pt}
    \label{fig:syntactic}
\end{figure*}

{\subsubsection{Syntactic Characteristics} 

\begin{itemize}[leftmargin=*]
    \item \textbf{Conditional Error} indicates there is an error within the `\code{if}' statement, causing the code to behave incorrectly. 
    \item \textbf{Loop Error} indicates there is an iteration mistake in the `\code{for}' or `\code{while}' loop, either through incorrect loop boundaries or mismanagement of loop variables.
    \item \textbf{Return Error} indicates the error is in a return statement that returns a wrong value or a value in the unexpected format.
    \item \textbf{Method Call Error} indicates the error is in a function call. It can be \textit{incorrect function name}, wrong arguments (\textit{incorrect function arguments}), or \textit{incorrect method call target}. 
    \item \textbf{Assignment Error} indicates the error is in an assignment statement. It can be an incorrect constant/variable name/comparison operator used in an assignment, leading to errors or unexpected behaviors in the code's execution.
    \item \textbf{Import Error} indicates the error is in an import statement.
    \item \textbf{Code Block Error} indicates multiple statements are incorrectly generated or omitted, leading to the task failure.
  
\end{itemize}
}

\smalltitle{Comparison between LLMs.} Fig.~\ref{fig:syntactic} shows the distribution of the 14 syntactic characteristics across the six LLMs. We observed similar distribution patterns as semantic characteristics. For all models, the top 3 error locations are either in entire code blocks (i.e., multiple statements in a sequence) or in an if statement. The fact that all LLMs struggle with generating entire code blocks correctly implies that many code generation errors are not small errors and require substantial efforts to fix, as investigated further in RQ2.

Compared with other models, the code generation errors from GPT-4 are more well-contained in a few types of code constructs. GPT-4 did not introduce any errors in method call expressions,  variable references, or constant values used in an assignment statement. By contrast, GPT-3.5 still hallucinates when generating method calls. Other models exhibited a more diverse set of error locations compared with GPT-4 and GPT-3.5.  Interestingly, CodeGen-16B and InCoder-1.3B have more cases of \textit{incorrect function name}, while GPT-3.5, SantaCoder, and StarCoder encounter \textit{incorrect function arguments} more frequently. This implies that during pre-training, CodeGen and InCoder are less effective in learning the mappings between task descriptions and which functions to use to achieve the tasks. One interesting direction to improve these models is to design pre-training tasks that predict function names and arguments to strengthen the model's memory of function usage.


\vspace{-5pt}
\begin{finding}
    \label{finding:3}
    More than 40\% of the syntactic characteristics made by six LLMs are \textit{missing/incorrect code block}. The studied LLMs also encountered a significant number of \textit{if error} and \textit{incorrect function name/argument}. 
\end{finding}
\vspace{-5pt}

\begin{figure*}[t]
     \centering
     \vspace{-10pt}
     \begin{subfigure}[b]{0.28\textwidth}
         \centering
         \includegraphics[width=1\linewidth]{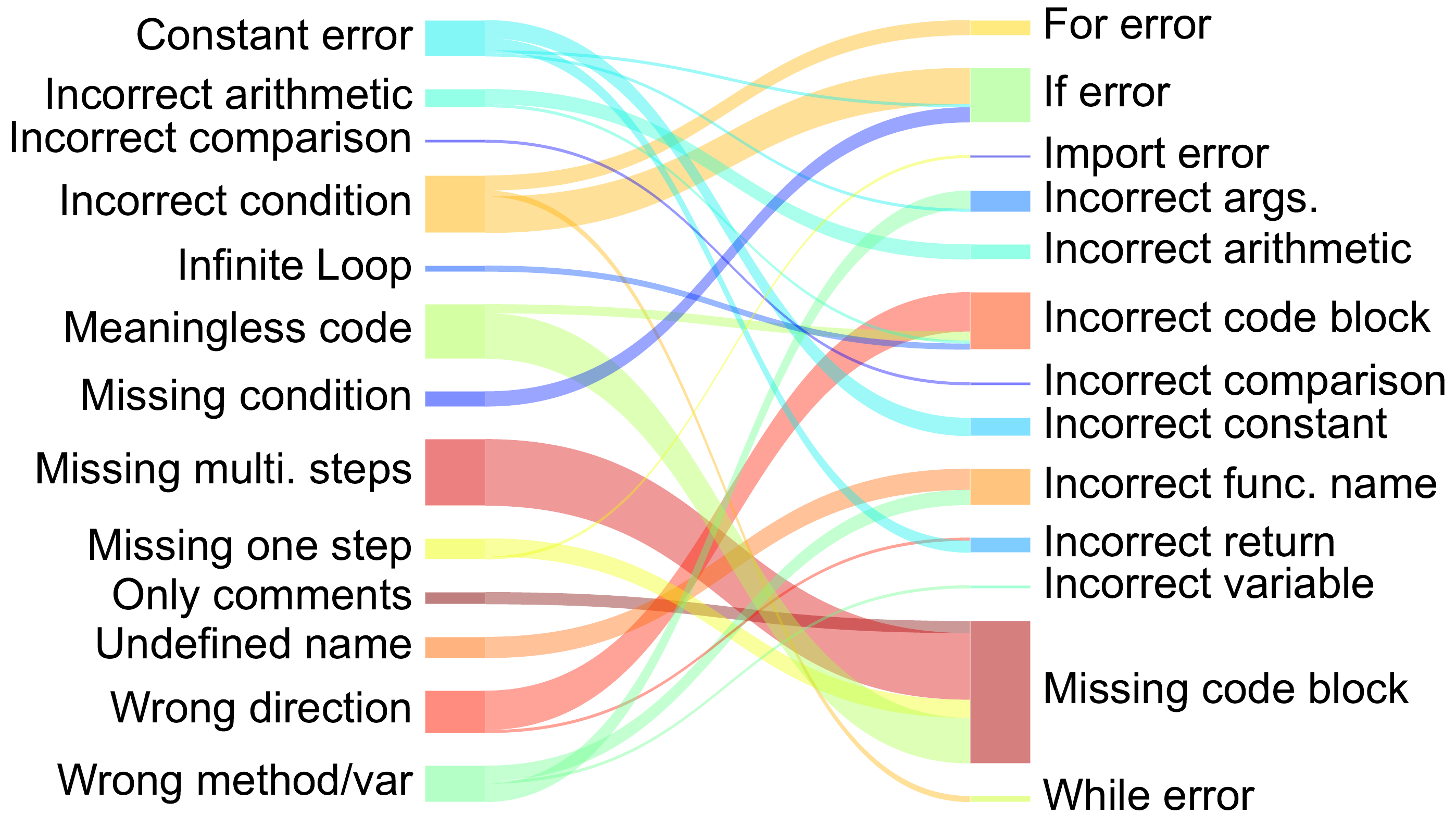}
         \vspace{-17pt}
        \caption{\responseref{}CodeGen-16B}
        \vspace{-8pt}
         \label{fig:codegen_sankey}
     \end{subfigure}%
     \hspace{4mm}
     \begin{subfigure}[b]{0.28\textwidth}
         \centering
         \includegraphics[width=1\linewidth]{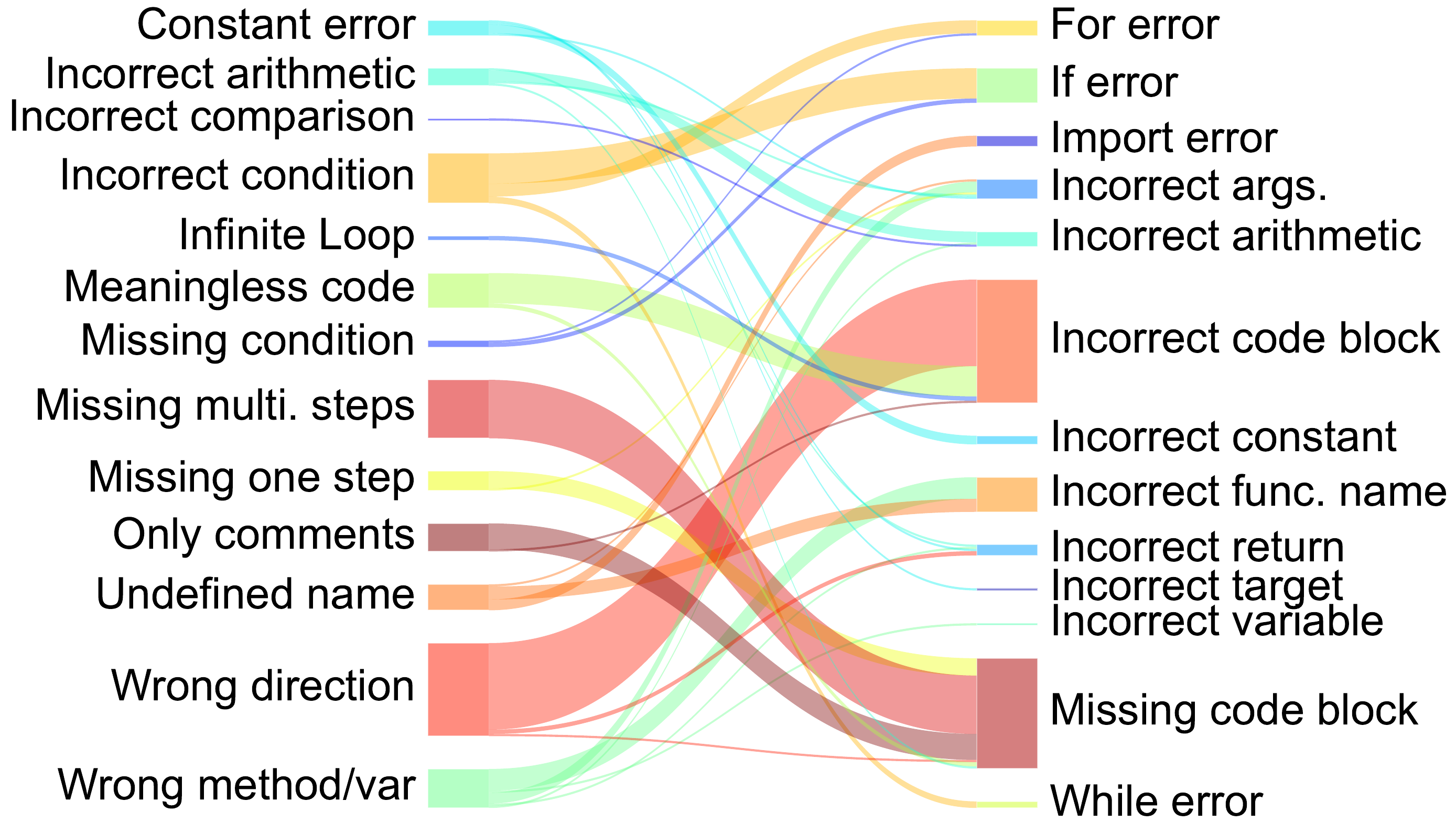}
         \vspace{-17pt}
        \caption{\responseref{}InCoder-1.3B}
        \vspace{-8pt}
         \label{fig:incoder_sankey}
     \end{subfigure}%
     \hspace{4mm}
     \begin{subfigure}[b]{0.28\textwidth}
         \centering
         \includegraphics[width=1\linewidth]{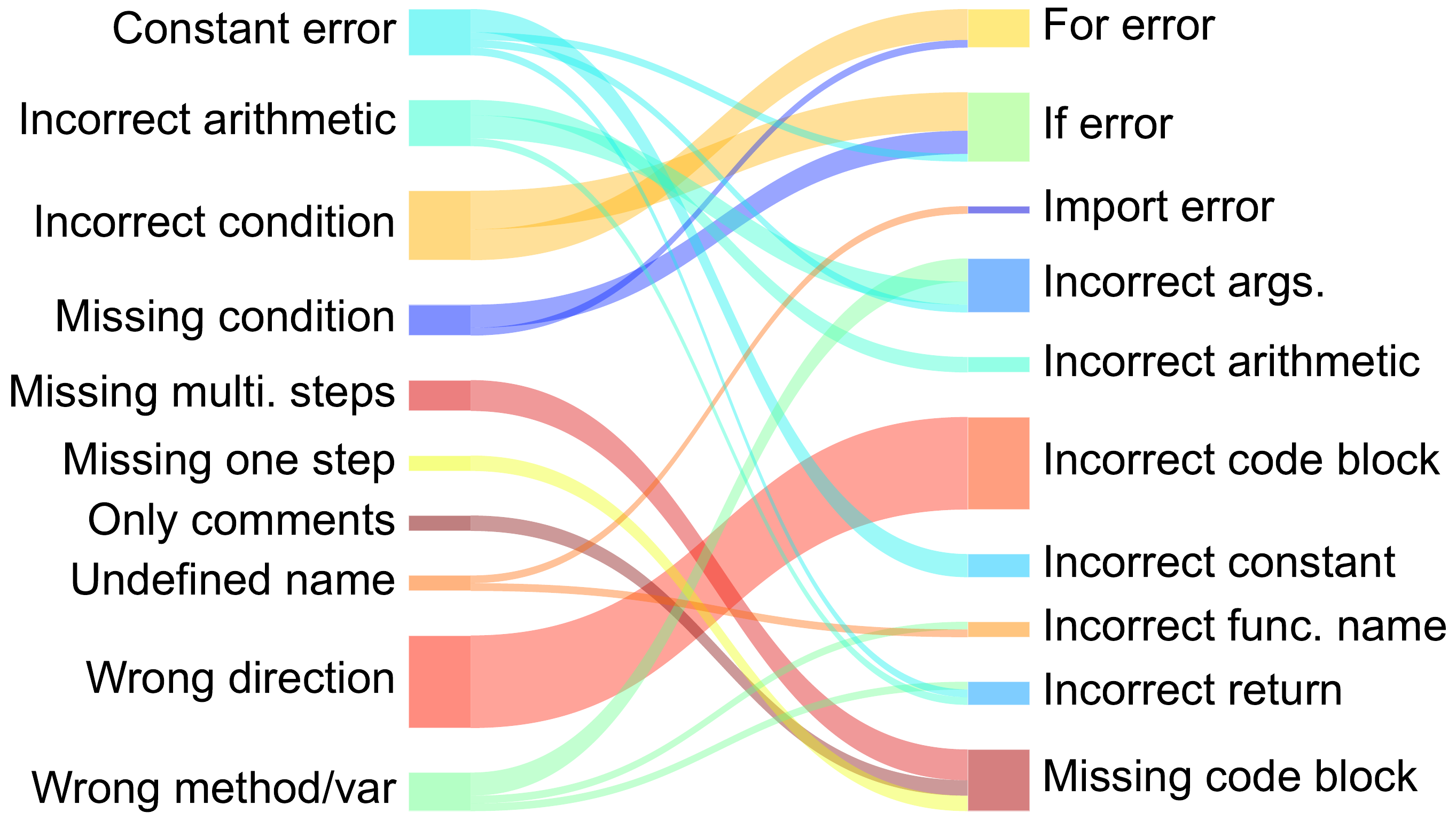}
         \vspace{-17pt}
         \caption{\responseref{}GPT-3.5}
         \vspace{-8pt}
         \label{fig:chatgpt_sankey}
     \end{subfigure}%
     \vfill
     \begin{subfigure}[b]{0.28\textwidth}
         \centering
         \vspace{10pt}
         \includegraphics[width=1\linewidth]{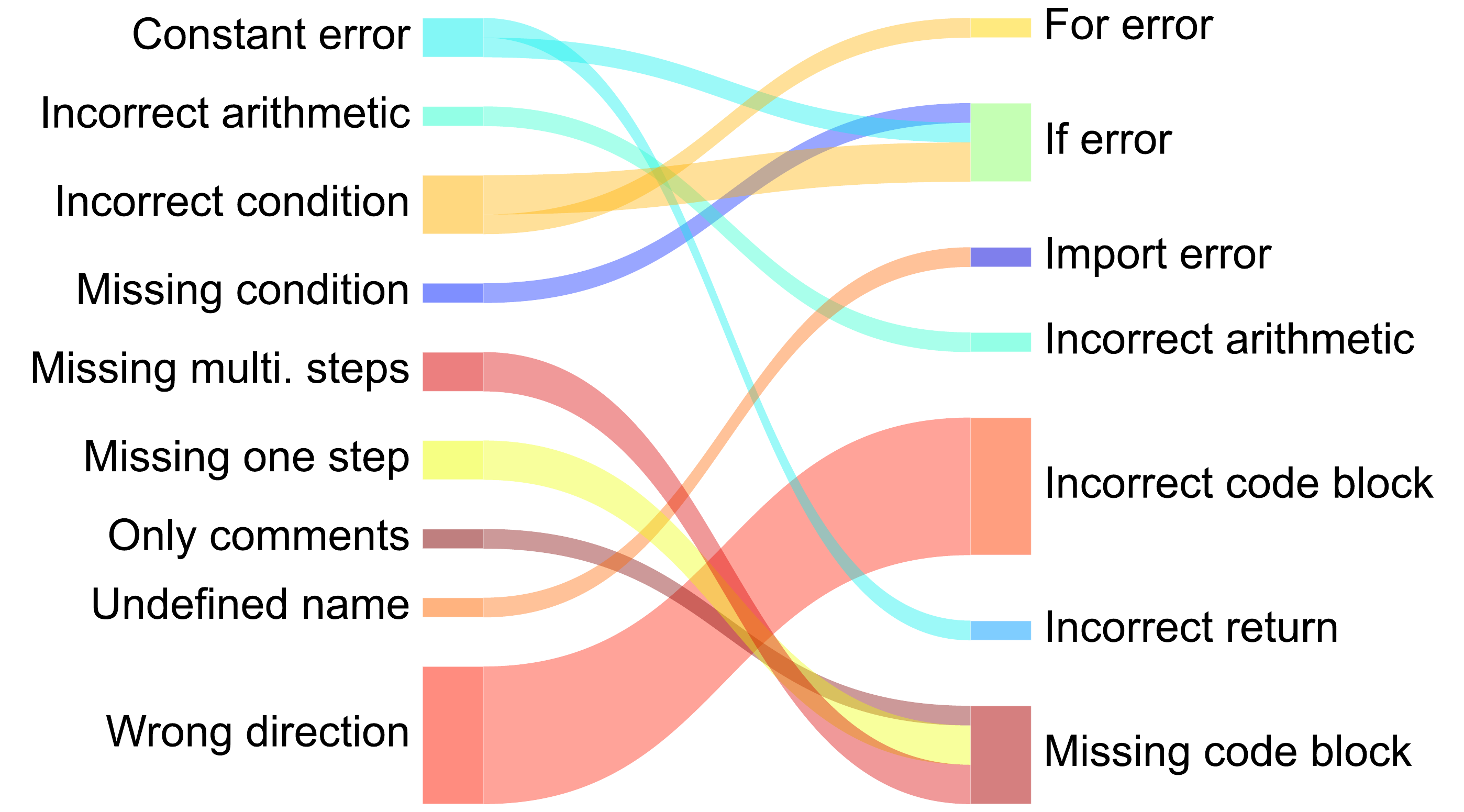}
         \vspace{-17pt}
         \caption{\responseref{}GPT-4}
         \vspace{-5pt}
         \label{fig:gpt4_sankey}
     \end{subfigure}%
     \hspace{3.5mm}
     \begin{subfigure}[b]{0.28\textwidth}
         \centering
         \vspace{10pt}
         \includegraphics[width=1\linewidth]{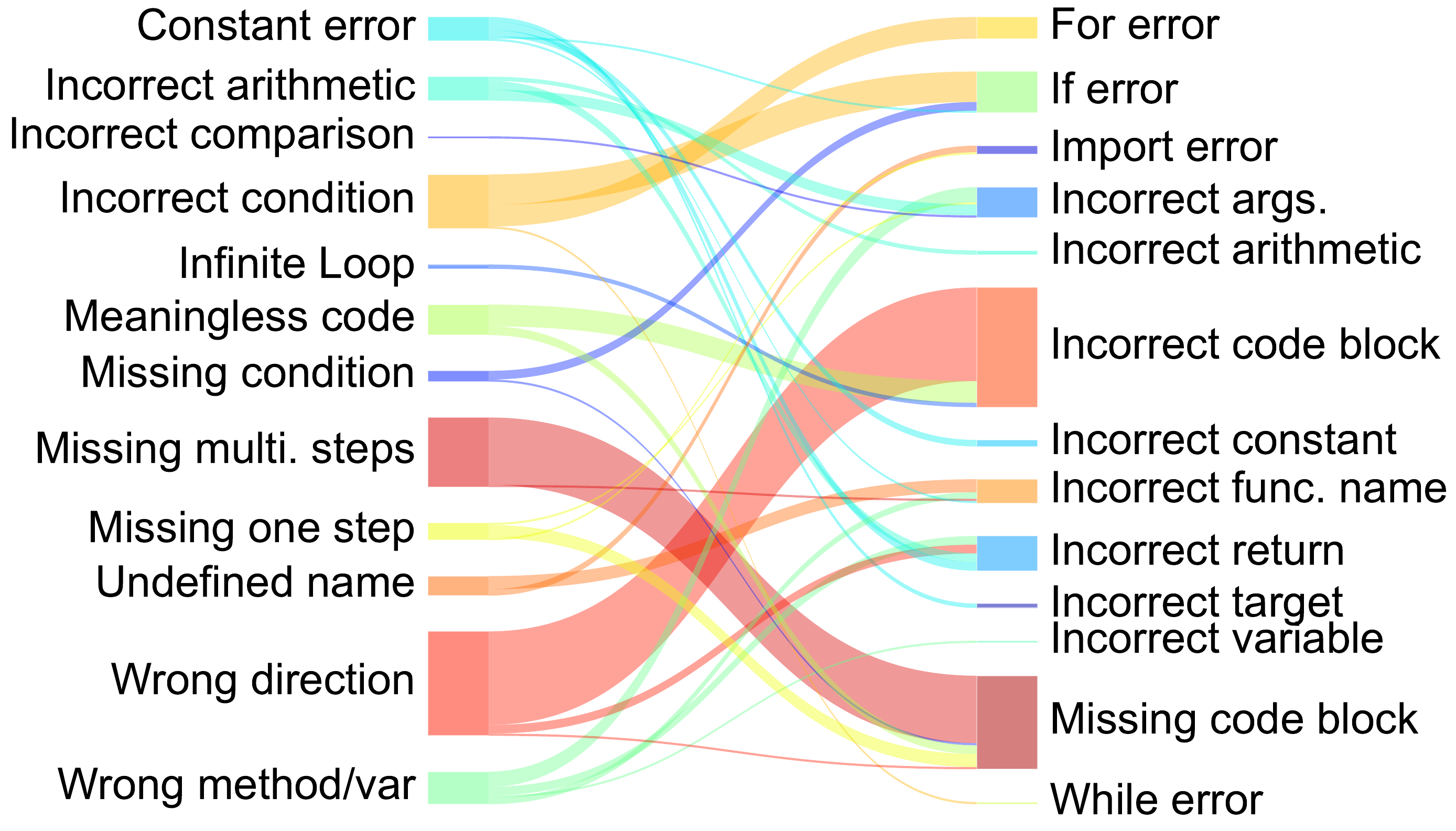}
         \vspace{-17pt}
         \caption{\responseref{}SantaCoder}
         \vspace{-5pt}
         \label{fig:santacoder_sankey}
     \end{subfigure}%
     \hspace{4mm}
     \begin{subfigure}[b]{0.28\textwidth}
         \centering
         \vspace{10pt}
         \includegraphics[width=1\linewidth]{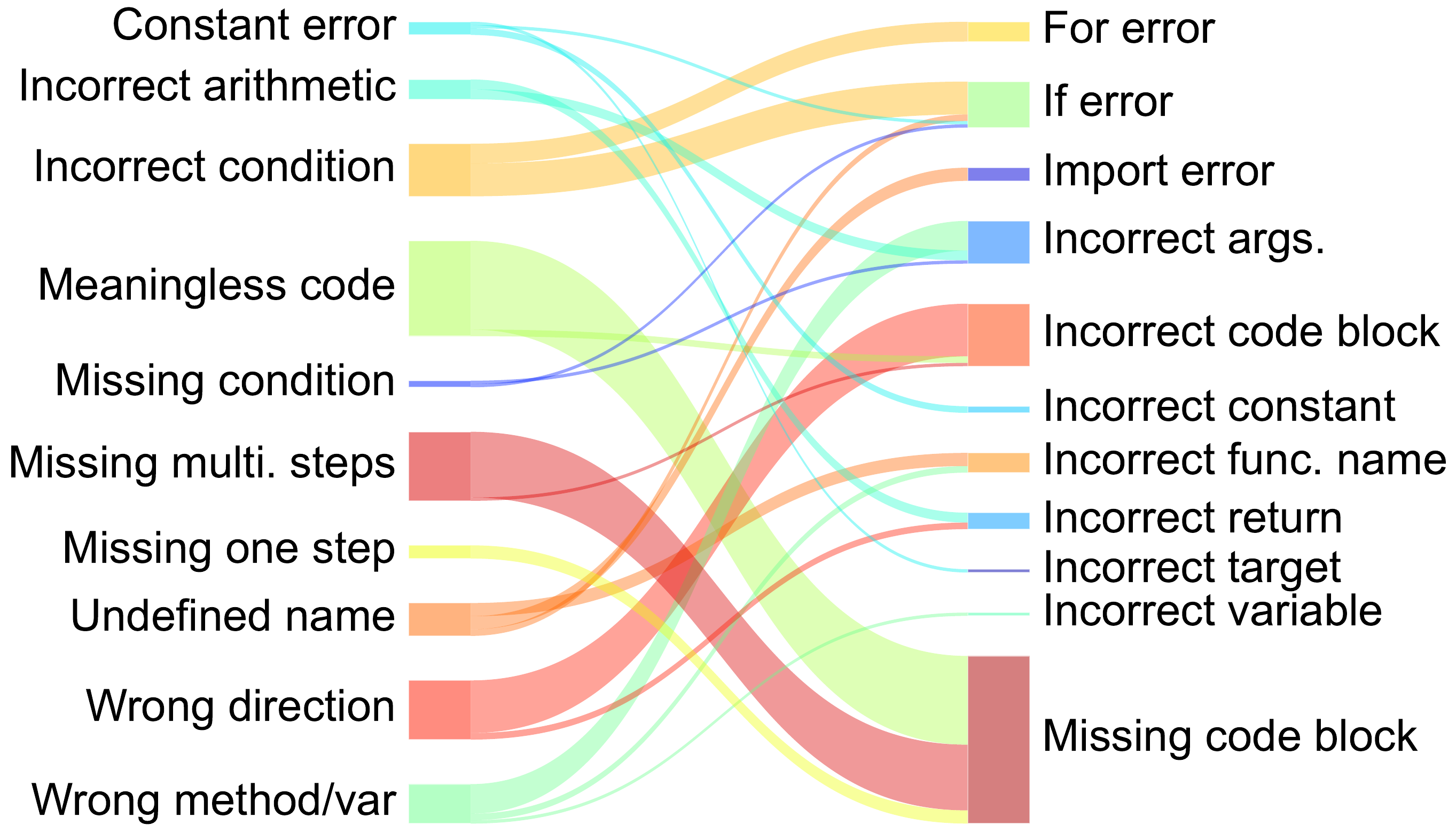}
         \vspace{-17pt}
         \caption{\responseref{}StarCoder}
         \vspace{-5pt}
         \label{fig:starcoder_sankey}
     \end{subfigure}%
    \caption{\responseref{} Mappings between semantic and syntactic error characteristics of code generation errors made by six LLMs.}
    \label{fig:sankey}
\end{figure*}

\begin{figure*}[t]
     \centering
     \vspace{-10pt}
     \begin{subfigure}[b]{0.23\textwidth}
         \centering
         \includegraphics[width=\linewidth]{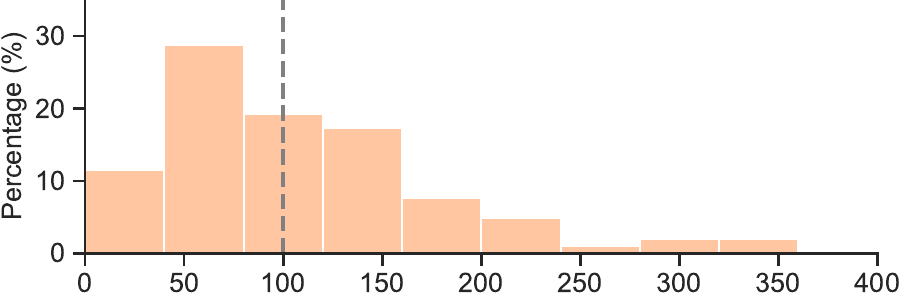}
         \vspace{-18pt}
         \caption{CodeGen-16B}
         \vspace{-13pt}
         \label{fig:codegen_leven}
     \end{subfigure}%
     \hspace{11mm}
     \begin{subfigure}[b]{0.23\textwidth}
         \centering
         \includegraphics[width=\linewidth]{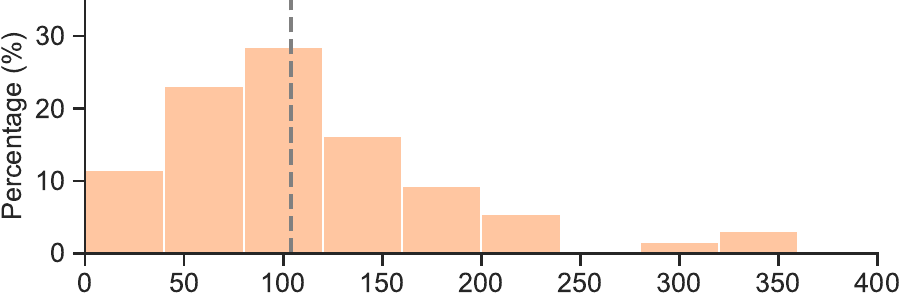}
         \vspace{-18pt}
         \caption{InCoder-1.3B}
         \vspace{-13pt}
         \label{fig:incoder_leven}
     \end{subfigure}%
     \hspace{11mm}
     \begin{subfigure}[b]{0.23\textwidth}
         \centering
         \includegraphics[width=\linewidth]{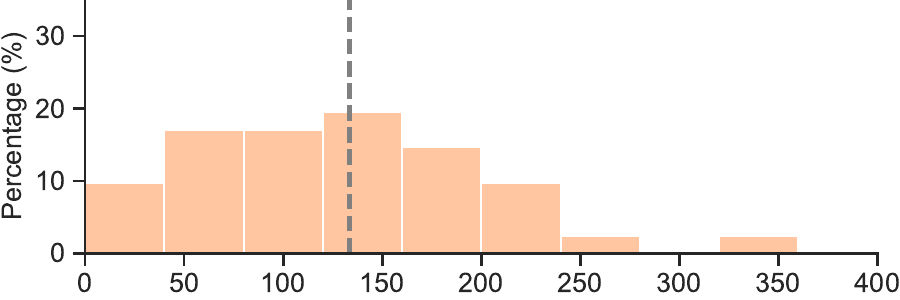}
         \vspace{-18pt}
         \caption{GPT-3.5}
         \vspace{-13pt}
         \label{fig:gpt_leven}
     \end{subfigure}%
      \vfill
     \begin{subfigure}[b]{0.23\textwidth}
         \centering
         \vspace{12pt}
         \includegraphics[width=\linewidth]{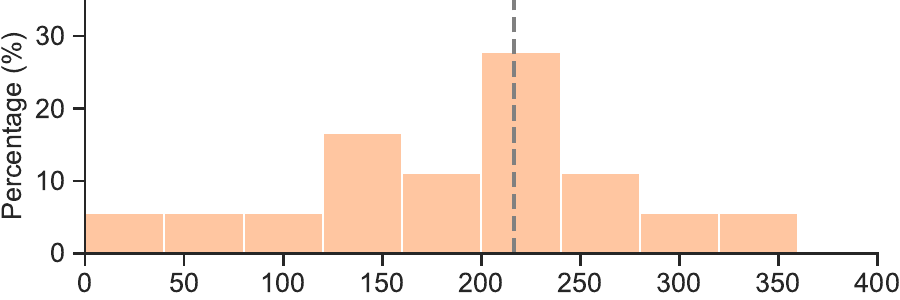}
         \vspace{-18pt}
         \caption{GPT-4}
         \vspace{-5pt}
         \label{fig:gpt4-leven}
     \end{subfigure}%
     \hspace{11mm}
     \begin{subfigure}[b]{0.23\textwidth}
         \centering
         \vspace{12pt}
         \includegraphics[width=\linewidth]{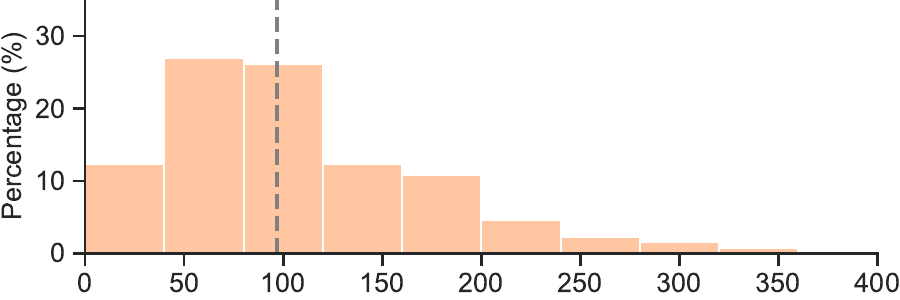}
         \vspace{-18pt}
         \caption{SantaCoder}
         \vspace{-5pt}
         \label{fig:santacoder_leven}
     \end{subfigure}%
     \hspace{11mm}
     \begin{subfigure}[b]{0.23\textwidth}
         \centering
         \includegraphics[width=\linewidth]{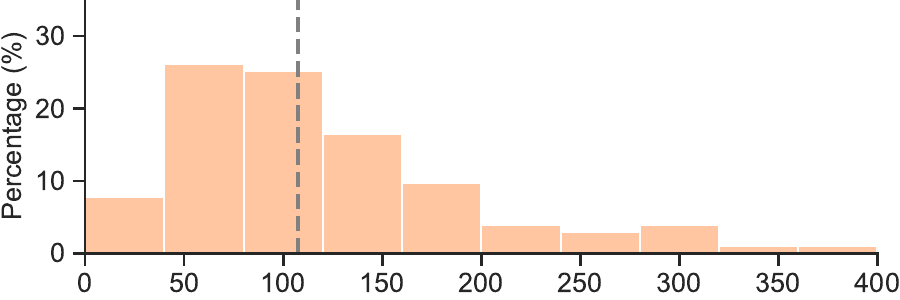}
         \vspace{-18pt}
         \caption{StarCoder}
         \vspace{-5pt}
         \label{fig:starcoder_leven}
     \end{subfigure}%
    \caption{Levenshtein distance between the incorrect code and correct code. The vertical dashed lines indicate the medians.}
    \vspace{-10pt}
    \label{fig:leven}
\end{figure*}

\begin{figure*}[t]
     \centering
     \begin{subfigure}[b]{0.23\textwidth}
         \centering
         \includegraphics[width=\linewidth]{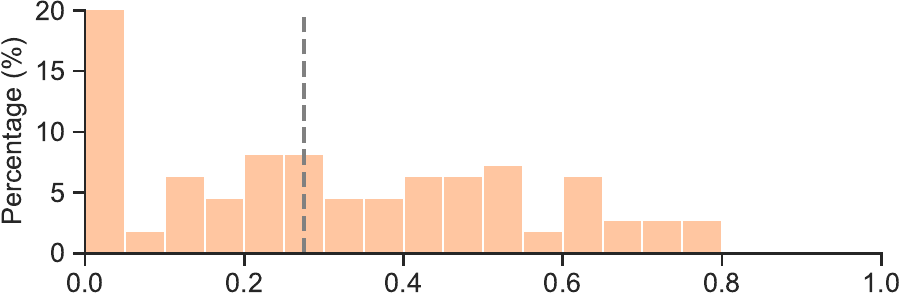}
         \vspace{-18pt}
         \caption{\responseref{}CodeGen-16B}
         \vspace{-13pt}
         \label{fig:codegen_codebert}
     \end{subfigure}%
     \hspace{10.5mm}
     \begin{subfigure}[b]{0.23\textwidth}
         \centering
         \includegraphics[width=\linewidth]{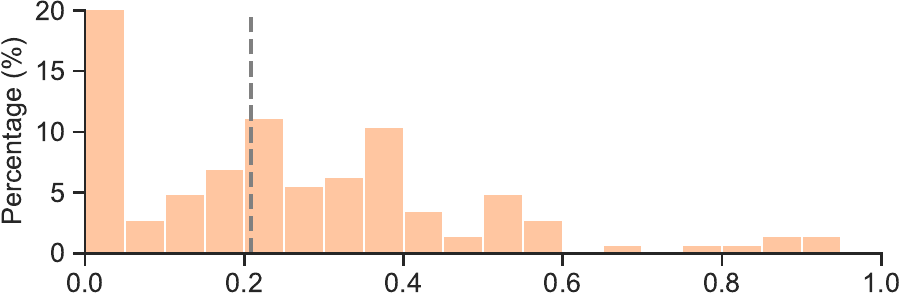}
         \vspace{-18pt}
         \caption{\responseref{}InCoder-1.3B}
         \vspace{-13pt}
         \label{fig:incoder_codebert}
     \end{subfigure}%
     \hspace{10.5mm}
     \begin{subfigure}[b]{0.23\textwidth}
         \centering
         \includegraphics[width=\linewidth]{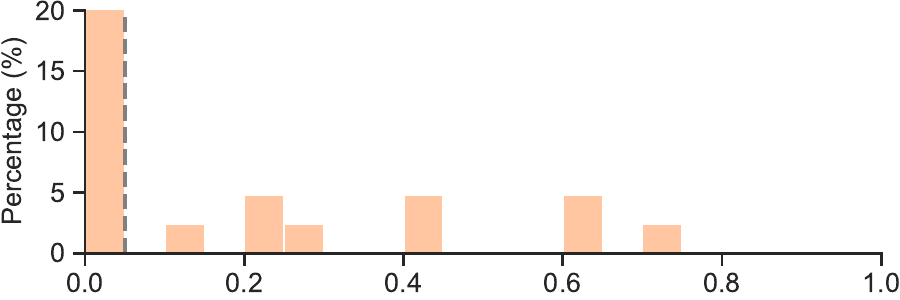}
         \vspace{-18pt}
         \caption{\responseref{}GPT-3.5}
         \vspace{-13pt}
         \label{fig:gpt_codebert}
     \end{subfigure}%
      \vfill
     \begin{subfigure}[b]{0.23\textwidth}
         \centering
         \vspace{12pt}
         \includegraphics[width=\linewidth]{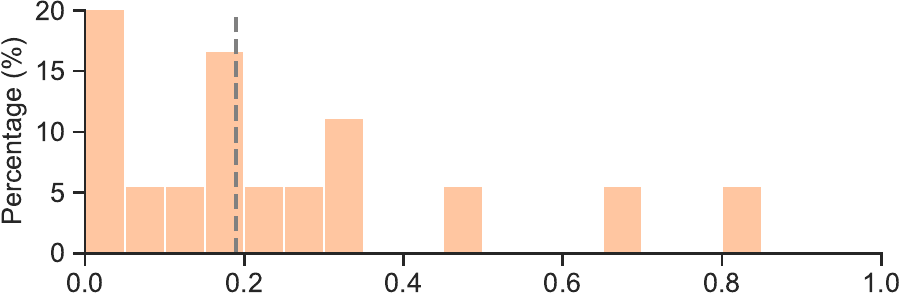}
         \vspace{-18pt}
         \caption{\responseref{}GPT-4}
         \vspace{-5pt}
         \label{fig:gpt4-codebert}
     \end{subfigure}%
     \hspace{10.5mm}
     \begin{subfigure}[b]{0.23\textwidth}
         \centering
         \vspace{12pt}
         \includegraphics[width=\linewidth]{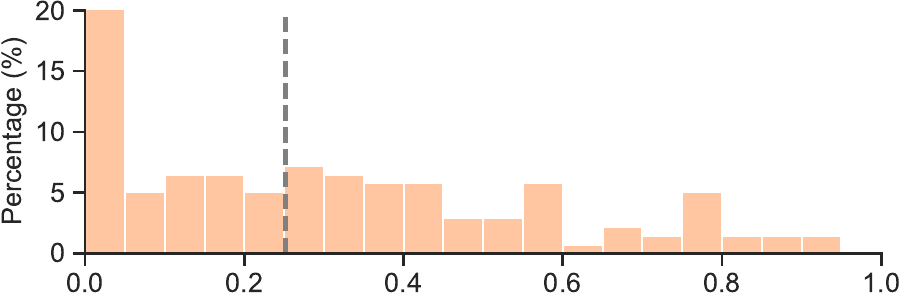}
         \vspace{-18pt}
         \caption{\responseref{}SantaCoder}
         \vspace{-5pt}
         \label{fig:santacoder_codebert}
     \end{subfigure}%
     \hspace{10.5mm}
     \begin{subfigure}[b]{0.23\textwidth}
         \centering
         \includegraphics[width=\linewidth]{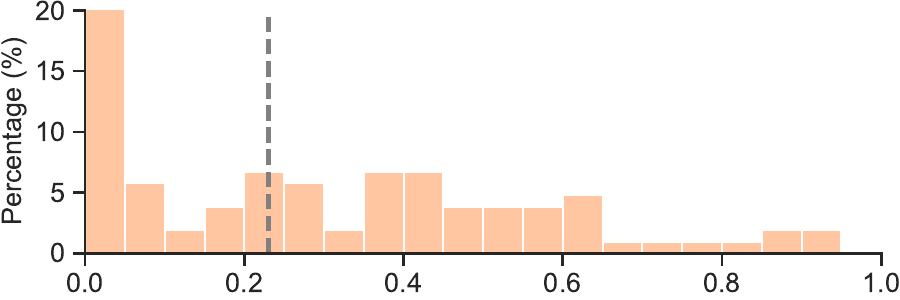}
         \vspace{-18pt}
         \caption{\responseref{}StarCoder}
         \vspace{-5pt}
         \label{fig:starcoder_codebert}
     \end{subfigure}%
    \caption{\responseref{} CodeBERTScore between the incorrect code and correct code. The vertical dashed lines indicate the medians.}
    \vspace{-20pt}
    \label{fig:codebert}
\end{figure*}

{\responseref

\subsubsection{Mappings Between Semantic and Syntactic Characteristics}

To further investigate the relationship between the semantic and syntactic characteristics of these code generation errors, we visualize their mappings as Sankey diagrams in Fig.~\ref{fig:sankey}. 
We observed that certain semantic characteristics are often paired with specific syntactic characteristics. For instance, \textit{wrong (logical) direction} typically corresponds to \textit{incorrect code block}. In other cases, a single syntactic characteristic can be associated with multiple semantic characteristics. For example, an \textit{incorrect function argument} might arise from \textit{constant value error}, \textit{incorrect arithmetic operation}, or \textit{wrong method/variable}. These results indicate that errors in similar locations may have various semantic root causes and thus require different kinds of fixes. 

For example, to fix SantaCoder's solution for Task 50 (Example~\ref{exp-sankey}), one only needs to replace the constant \texttt{97} with \texttt{26}. By contrast, fixing the incorrect function argument in CodeGen-16B’s solution for Task 149 is more challenging, as it requires generating an additional comparator function as the argument of the lambda function (Example~\ref{exp-sankey}).

}

{
\setlength{\fboxsep}{0pt}
\begin{lstlisting}[language=Python, escapechar=!, numberstyle=\tiny\color{lightgray}]
# [Task 50] Decode the string that is encoded by shifting
# 5 characters in the alphabet
def decode_shift(s):
    return "".join([chr(((ord(ch) - 5 +!\colorbox{lightred}{ 97}!) % 26) + ord("a")) for ch in s])
\end{lstlisting}
\vspace{-2pt}
\begin{lstlisting}[language=Python, escapechar=!, numberstyle=\tiny\color{lightgray}, caption=\responseline{\textit{Incorrect function argument}}, label=exp-sankey]
# [Task 149] Delete the strings that have odd lengths.
# Return a sorted list ascending by length of each word.
# Sort alphabetically for words of the same length.
def sorted_list_sum(lst)
    return sorted(list(filter(lambda x: len(x)%2==0, lst)), !\colorbox{lightred}{key=\codeoperation{len}}!)
\end{lstlisting}
\vspace{-3pt}
}

{\responseref

We also found that one kind of semantic characteristic, such as \textit{constant value error}, may occur in different kinds of locations, such as if statements and function calls. Precisely locating such errors can, therefore, be challenging. For instance, in Task 91 (Example~\ref{exp-sankey2}), GPT-4 should execute \texttt{startswith('I ')} instead of \texttt{startswith('I')} since a boredom is a sentence that starts with the {word} ``I'' rather than the {character} ``I.'' While the fix itself is small, it requires a deep understanding of the task to derive the fix.

}

\begin{lstlisting}[language=Python, escapechar=^, numberstyle=\tiny\color{lightgray}, caption=\responseline{\textit{Constant value error} by GPT-4}, label=exp-sankey2]
# [Task 91] Count the number of boredoms. A boredom is a 
# sentence that starts with the word "I". Sentences are 
# delimited by '.', '?' or '!'.
def is_bored(S):
    sentences = [s.strip() for s in re.split('[.!?]', S)]
    return sum(1 for s in sentences if s.startswith("I"))
\end{lstlisting}

{\responseref

\subsubsection{The Impact of Training Data}

Regarding semantic characteristics, we found that GPT-3.5 and GPT-4 did not generate any \textit{meaningless code snippets}, unlike the other four LLMs. This might be due to their training on significantly larger datasets, enhancing their ability to avoid meaningless outputs. In terms of syntactic characteristics, we observed that CodeGen-16B produced a smaller proportion of \textit{incorrect code blocks} compared to the other models. 
This could be attributed to its specialized training with Python code data alone, which possibly contributed to reducing the generation of large chunks of syntactically incorrect code. In contrast, the other LLMs were trained on code corpora of multiple programming languages.

}

\subsection{RQ2: Repair Effort for Code Generation Errors}
\label{subsec:rq4}

Fig.~\ref{fig:leven} shows the distribution of Levenshtein distances. All models exhibit a wide range of Levenshtein distances for incorrect code, with median distances around or greater than 100. Notably, 84.21\% of the incorrectly generated code has Levenshtein distance scores above 50 edits, with 52.63\% of them requiring more than 200 edits. \responseline{The results of Jaccard similarity show similar trends as the Levenshtein distances. Due to the page limit, we refer readers to our GitHub repository~\cite{ourwebsite} for the detailed results of Jaccard similarity.}

\responseline{As shown in Fig.~\ref{fig:codebert}, the median CodeBERTScore values are approximately 0.2 for the six LLMs, except for GPT-3.5, which has a median similarity of 0.05. Additionally, a large portion of CodeBERTScore values across all six LLMs falls below 0.1. According to~\cite{zhou2023codebertscore}, a CodeBERTScore value of 0 indicates that two code snippets are unrelated, while a value of 1 indicates that the snippets are exactly the same. These results indicate that the majority of the incorrect solutions deviate significantly in semantics from the correct solutions. Overall, both low textual and semantic similarities suggest that the incorrect code generated by LLMs often exhibits big differences from the correct solutions, not just minor errors.} Such observations are also aligned with our findings in RQ1---LLMs tend to make non-trivial errors such as \textit{missing multiple steps} and \textit{wrong logical direction}. Addressing these issues would require substantial edits. For instance, the median number of edits to fix errors of \textit{missing multiple steps} is 108 edits. 

Interestingly, GPT-3.5 and GPT-4, despite having high performance (Table~\ref{table:model_table}), exhibit larger deviations when generating incorrect code, with greater median Levenshtein distances compared to other models. This suggests that though GPT-3.5 and GPT-4 are more accurate in general, when they make mistakes, the mistakes are likely to cause a larger deviation (e.g., a large \textit{incorrect code block}) from the correct solution and thus require more edits to fix.

\begin{figure}[t]
    \vspace{-8pt}
    \centering
    \includegraphics[width=0.9\linewidth]{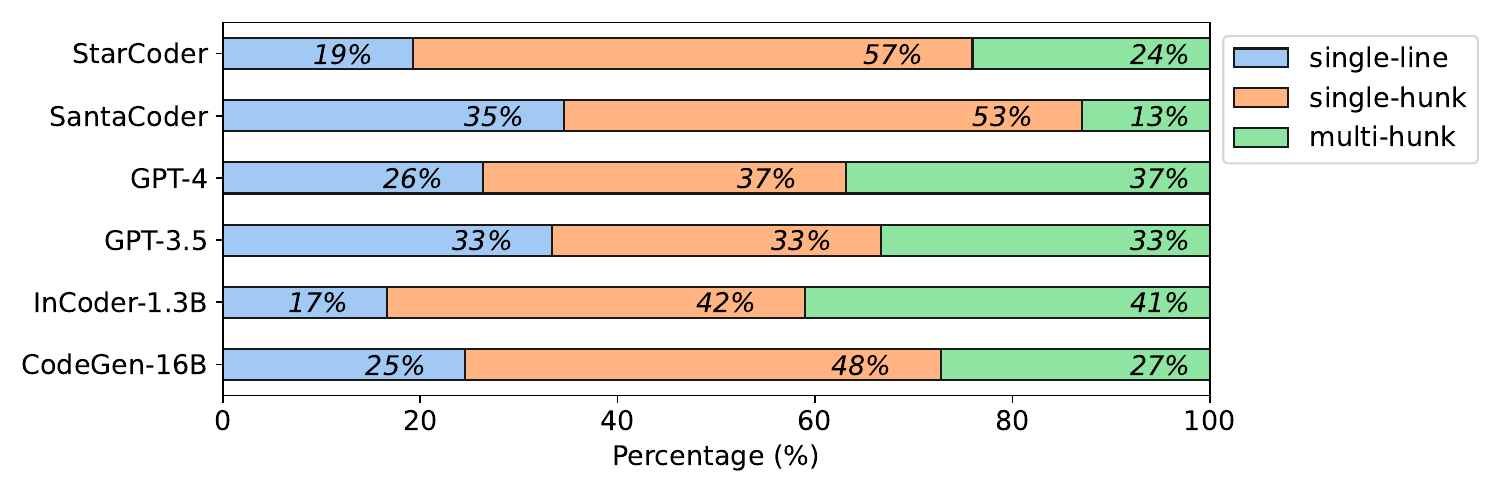}
    \vspace{-8pt}
    \caption{Distribution of \textit{single-line}, \textit{single-hunk}, and \textit{multi-hunk} faults in different LLMs}
    \vspace{-17pt}
    \label{fig:line_and_hunk}
\end{figure}

Following the automated program repair literature~\cite{saha2019harnessing}, we also classify all incorrect code snippets into three categories based on the effort required to fix them: (1) single-line errors, (2) single-hunk errors, and (3) multi-hunk errors. Specifically, a ``hunk'' refers to several contiguous lines in a program. Fig.~\ref{fig:line_and_hunk} shows the distribution. Overall, the majority of errors are single-hunk or multi-hunk errors, which require substantial effort to repair compared with single-line errors. Compared with other LLMs, GPT-3.5 and GPT-4 exhibit a more balanced distribution among the three categories. SantaCoder made the most single-line errors (35\%). StarCoder made the most single-hunk errors (57\%). Notably, the least accurate model, InCoder-1.3B, made the most multi-hunk errors (41\%). 

\vspace{-5pt}
\begin{finding}
    \label{finding:4}
    The majority of incorrect code solutions generated by the six LLMs deviate significantly from the correct code. This implies that fixing LLM-generated code requires non-trivial efforts. 
\end{finding}
\vspace{-5pt}

\subsection{RQ3: Impact of Task Complexity}
\label{subsec:rq5}

To the best of our knowledge, there is no established metric to measure task complexity. In this study, we used the length of the task description (i.e., the number of words in the prompt) and the length of the correct solution in terms of lines of code (LOC) as proxy metrics for task complexity. It is a pragmatic choice to enable objective measurements, but we acknowledge its limitation and discuss the potential threats to validity in Sec.~\ref{sec:threats}. \responseline{To avoid inflating the prompt length, we removed all test cases from the prompt. We also noticed that 20 ground-truth solutions only contain one LOC but with complex constructs such as lambda expressions. Therefore, we used the number of abstract syntax tree (AST) nodes as an alternative metric to LOC to verify its validity.}

\begin{figure}[t]
    \centering
    \vspace{-7pt}
    \begin{subfigure}[b]{0.78\linewidth}
         \centering
         \includegraphics[width=\linewidth]{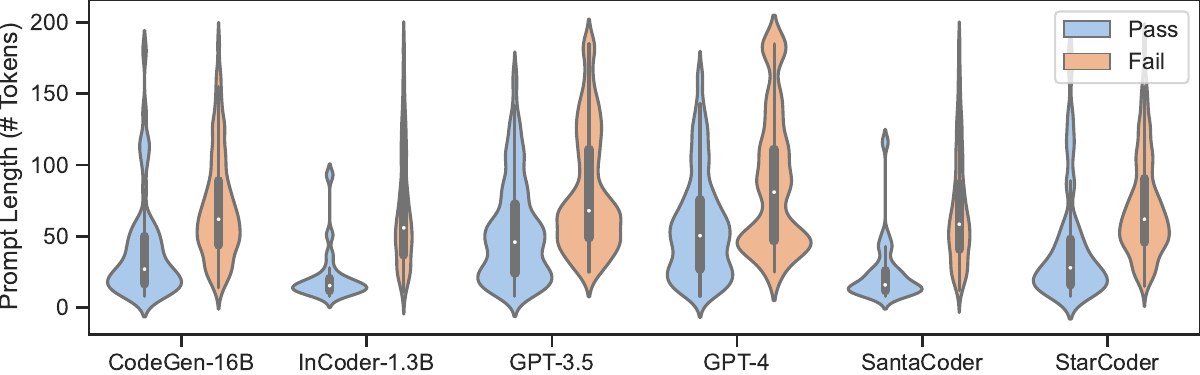}
         \vspace{-17pt}
         \caption{\responseline{Prompt length}}
    \label{fig:prompt_length_pass}
    \end{subfigure}%
    \vspace{1pt}
    \begin{subfigure}[b]{0.78\linewidth}
         \centering
         \includegraphics[width=\linewidth]{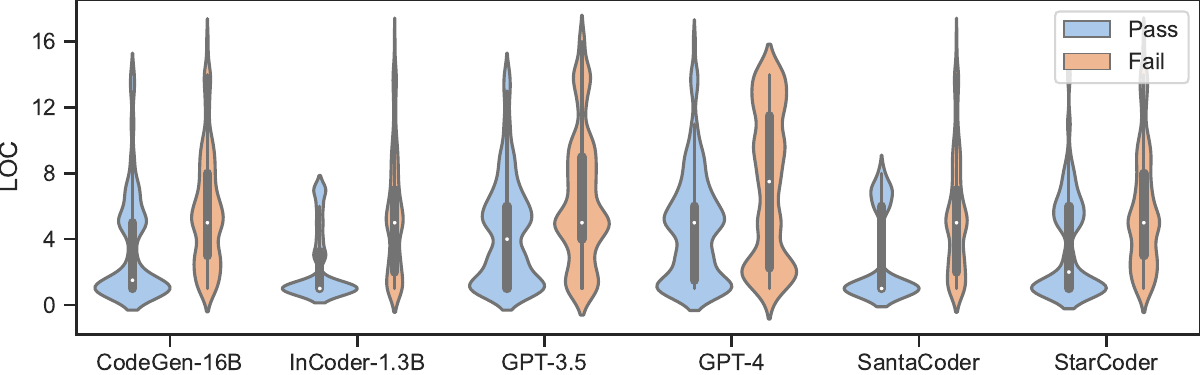}
         \vspace{-17pt}
         \caption{LOC of correct solutions}
    \label{fig:loc_pass}
     \end{subfigure}%
     \vspace{1pt}
    \begin{subfigure}[b]{0.78\linewidth}
         \centering
         \includegraphics[width=\linewidth]{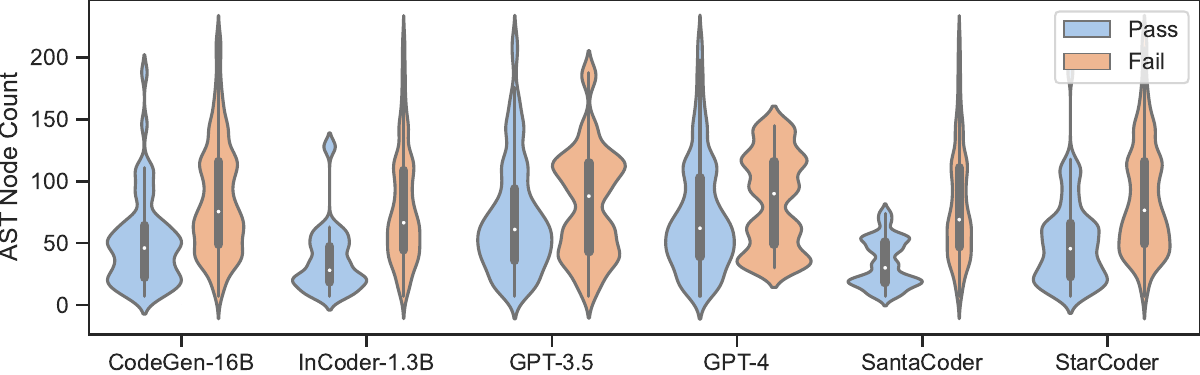}
         \vspace{-17pt}
         \caption{\responseref{} Number of AST nodes of correct solutions}
    \label{fig:ast_node}
     \end{subfigure}%
     \vspace{-5pt}
    \caption{Distribution of passed/failed tasks' complexity}
    \vspace{-18pt}
    \label{fig:task_complexity}
\end{figure}

We investigated whether there was a significant difference between the complexity of successfully solved tasks and the complexity of failed tasks. We ran the Mann-Whitney $U$-test to examine the statistical difference in task complexity for each LLM. Fig.~\ref{fig:task_complexity} shows the results. We observed a statistically significant difference in both prompt length and LOC of correct solutions among all six LLMs. \responseline{Specifically, the p-values for prompt length are 4e-9, 5e-9, 2e-5, 2e-10, 2e-9, and 5e-4; and the effect sizes are 0.82, 1.38, 0.71, 0.77, 1.32, and 0.75 for the six models in the order listed in Fig.~\ref{fig:prompt_length_pass}.} The p-values for LOC of correct solutions are 3e-6, 4e-5, 6e-3, 1e-2, 1e-5, and 1e-4; and the effect sizes are 0.72, 0.77, 0.63, 0.65, 0.69, and 0.73, respectively. \responseline{The number of AST nodes shows a similar distribution as LOC (Fig.~\ref{fig:ast_node}), where the correct solutions of successfully solved tasks have significantly more AST nodes. The p-values are 2e-7, 4e-5, 6e-3, 1e-2, 1e-5, and 1e-4; the effect sizes are 0.65, 0.88, 0.30, 0.43, 0.64, and 0.74. We further calculated the Pearson correlation coefficient between the number of AST nodes and LOC, which shows a statistically significant correlation (r=0.864, p=4e-50). This shows the consistency of our findings based on different metrics.} 

We further dig into each LLM's failed tasks with a prompt longer than 150 words. Based on the semantic taxonomy described in Sec.~\ref{subsec:rq1}, 
a large portion (64.0\%) of these failed tasks are \textit{Garbage Code} (i.e., unnecessary or irrelevant code that does not contribute to the intended functionality). 
For instance, Task 129's prompt includes 249 words, which requires returning a path with a length of k in a given grid that has the minimum sum of values alongside this path. InCoder-1.3B failed to understand the complex task requirement and only generated a sequence of meaningless appending operations \responseline{(Example~\ref{exp-long-prompt})}.

\begin{lstlisting}[language=Python, escapechar=!, numberstyle=\tiny\color{lightgray}, caption=\textit{Meaningless code snippet} by InCoder-1.3B, label=exp-long-prompt]
# [Task 129] Given a grid with N rows and N columns (N >= 
# 2) and a positive k, each cell of ... (230 words left)
def minPath(grid, k):
    ans. visited = [], {}
    visited[grid[0][0]] = True
    ans.append(grid[0][0])
    ans.append(grid[0][1])
    ...
\end{lstlisting}


We also observed similar patterns in the LOC metric. Specifically, we looked into programming tasks whose ground-truth solution has more than 12 lines of code. We found that 55.9\% of these failed tasks are \textit{Garbage Code}. For example, Task 105 requires three steps: 1) sorting an array of integers between 1 and 9 inclusive, 2) reversing the sorted result, and 3) replacing each digit with its corresponding name as a string. The correct solution has 24 lines of code. However, CodeGen-16B failed to understand the task requirements and returned only an empty array \responseline{(Example~\ref{exp-long-prompt-empty-code})}.

\begin{lstlisting}[language=Python, escapechar=!, numberstyle=\tiny\color{lightgray}, caption=\textit{Meaningless code snippet} by CodeGen-16B, label=exp-long-prompt-empty-code]
# [Task 105] Given an array of integers, sort the integers 
# that are between 1 and 9 inclusive, reverse the 
# resulting array, and then replace each digit by its 
# corresponding name "One", "Two" ... (144 words left)
def by_length(arr): return []
\end{lstlisting}
\vspace{-5pt}

\begin{finding}
    \label{finding:long_prompt_garbage}
    %
    We observed a significant performance gap between short, simple problems and long, complex ones. When the task prompt exceeded 150 words or the correct solution required more than 12 lines of code (LOC), about 60\% outputs are \textit{Garbage Code}. This highlights the research opportunity of estimating the upper bound of the task complexity that code generation LLMs can handle.
\end{finding}
\vspace{-9pt}

\subsection{RQ4: Impact of Test Pass Rate}
\label{subsec:rq7}

Since HumanEval~\cite{chen2021evaluating} provides test cases for each task, we are interested in whether completely failed code (i.e., failing all test cases) has different error characteristic patterns compared with partially failed code (i.e., failing a subset of test cases). Note that we excluded the 19 cases identified in Sec.~\ref{sec:collection} that passed all tests but are not equivalent to the ground truths.

\begin{figure}[t]
    \centering
    \vspace{-15pt}
    \includegraphics[width=0.67\linewidth]{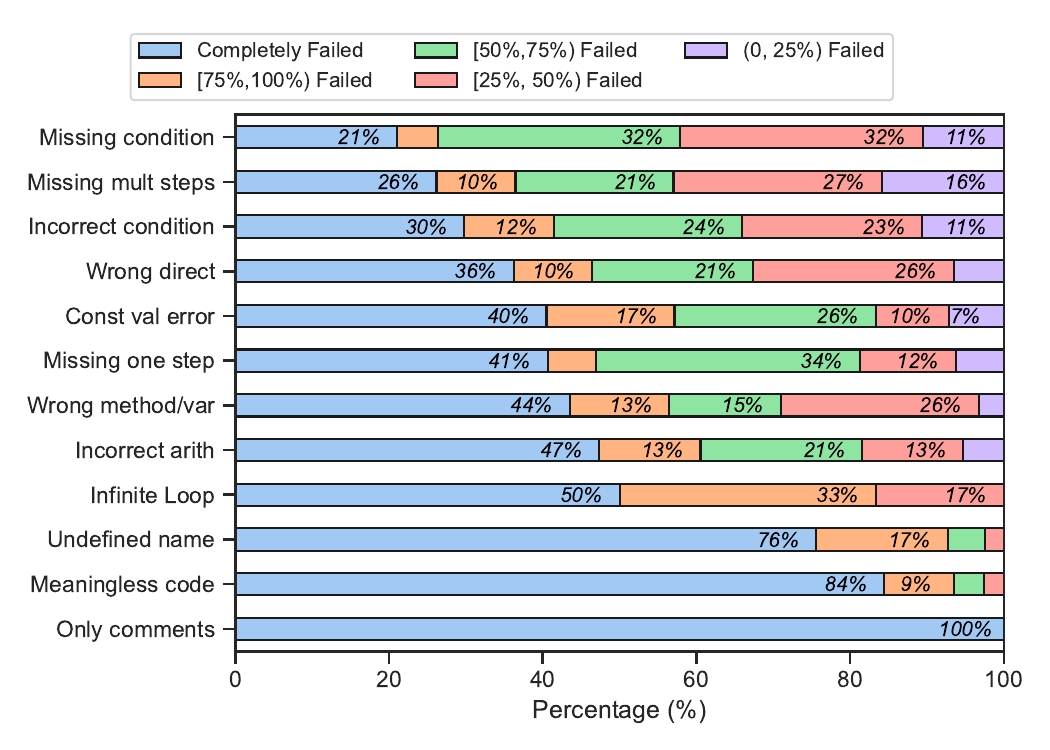}
    \vspace{-9pt}
    \caption{Semantic characteristics vs. test pass rates}
    \vspace{-15pt}
    \label{fig:test-pass-semantic}
\end{figure}

Fig.~\ref{fig:test-pass-semantic} shows the test pass rate distribution of incorrect solutions with different semantic characteristics. {\em Only comments} and {\em meaningless code} are the top two characteristics that lead to complete failures. Though {\em undefined name} sounds like a small error, the majority of code with {\em undefined names} also leads to complete test failures due to runtime crashes. Surprisingly, while some error characteristics, such as {\em missing multiple steps}, sound severe by definition, they do not often lead to complete failures. After digging into some instances, we noticed that this was because LLMs did not completely misunderstand the task description, and the generated code could still pass some weak test cases. 

For instance, Task 125 requires the program to split on commas if there is no space in the string. If there is also no comma in the string, the program should perform other operations. However, the code generated by CodeGen-16B only split the input string into spaces while missing the remaining steps. As a result, it can only pass the test cases that include spaces, leading to a partially failed task \responseline{(Example~\ref{exp-multi-step-partial})}.

{
\setlength\fboxsep{0pt}
\begin{lstlisting}[language=Python, escapechar=^, numberstyle=\tiny\color{lightgray}, caption=\textit{Missing multiple steps} by CodeGen-16B, label=exp-multi-step-partial]
# [Task 125] Given a string, return words split on space;
# if no space, split on ','; if no ',', return the number
# of lower-case letters with odd order in the alphabet.
def split_words(txt): return txt.split()
\end{lstlisting}
\vspace{-5pt}
}

Another surprising finding is that code with the wrong direction can pass some test cases accidentally. For example, in Task 75, the code generated by InCoder-1.3B does not follow the task instructions \responseline{(Example~\ref{exp-partial-test})}. However, it can pass a few test cases, such as 5, 10, and 30. One plausible reason is that since the prompt includes test cases, LLMs may have memorized a superficial correlation between test cases and some other irrelevant solutions that pass those test cases.

\begin{lstlisting}[language=Python, escapechar=!, numberstyle=\tiny\color{lightgray}, caption=\textit{Wrong (logical) direction} by InCoder-1.3B, label=exp-partial-test]
# [Task 75] Return true if the given number is the 
# multiplication of 3 prime numbers and false otherwise.
def is_multiply_prime(a): return a < 100 and a % 3 == 0
\end{lstlisting}
\vspace{-8pt}

\vspace{-5pt}
\begin{finding}
    \label{finding:9}
    When LLM-generated code only passes a subset of the given test cases, it is more likely to contain errors with \textit{missing multiple steps} and \textit{incorrect condition}. By contrast, completely failed code (no test cases passed) have \textit{meaningless code snippet} more frequently.
\end{finding}
\vspace{-5pt}

\begin{figure}[t]
    \centering
    \vspace{-15pt}
    \includegraphics[width=0.671\linewidth]{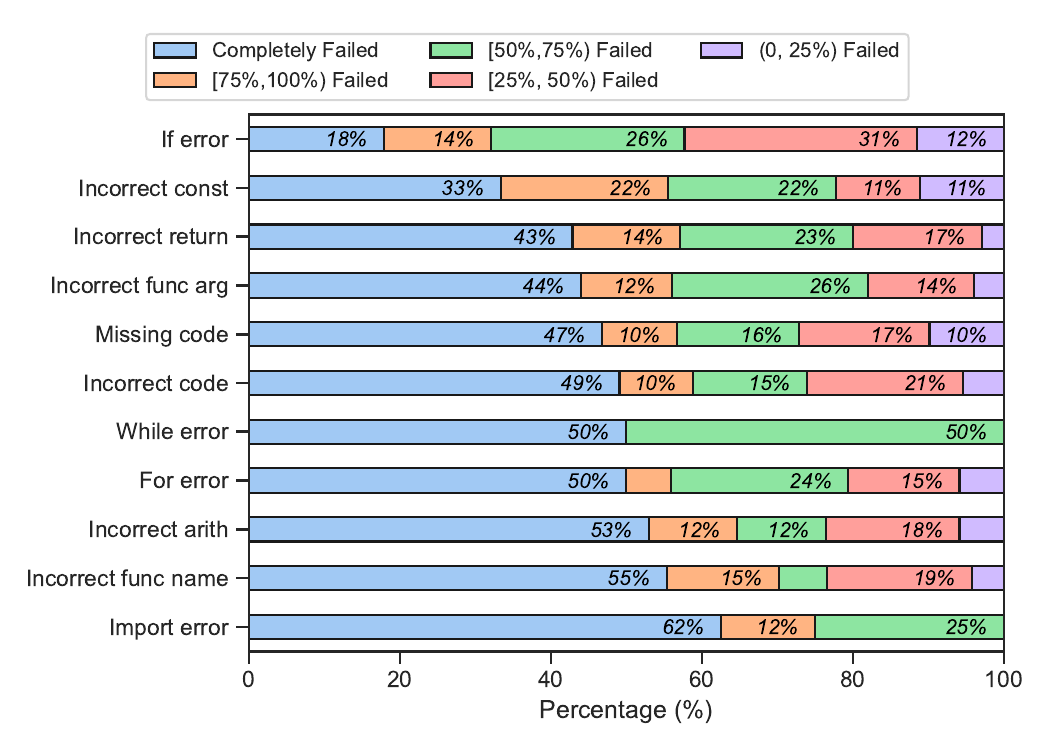}
    \vspace{-9pt}
    \caption{Syntactic characteristics vs. test pass rates}
    \vspace{-20pt}
    \label{fig:test-pass-syntactic}
\end{figure}

In terms of syntactic characteristics (Fig.~\ref{fig:test-pass-syntactic}), we observed an obvious difference on \textit{if error}, where the majority of code with this error led to partial failure (82\% v.s. 18\%). This is because when encountering \textit{if error}, LLMs were more likely to only misinterpret a specific condition requirement while generating correct code for the other parts. For instance, in Task 0, SantaCoder only considered the adjunct elements in the given list \responseline{(Example~\ref{exp-8})}. As a result, it failed to pass the test case where the two closet elements were not adjunct.

{\setlength\fboxsep{0pt}
\begin{lstlisting}[language=Python, escapechar=!, numberstyle=\tiny\color{lightgray}, caption=An \textit{if error} example by SantaCoder, label=exp-8]
# [Task 0] Check if in the given list of numbers, are any 
# two numbers closer to each other than given threshold.
def has_close_elements(numbers, threshold):
    if len(numbers) < 2: return False
    for i in range(len(numbers) - 1):
        if abs(!\colorbox{lightred}{numbers[i] - numbers[i+1]}!) > threshold: 
            return True
    return False
\end{lstlisting}
}
\vspace{-13pt}

\begin{finding}
    \label{finding:10}
    For syntactic characteristics, LLMs made more \textit{if error} in partially failed tasks (82\%) than in fully failed ones (18\%). These subtle error characteristics in partially failed tasks, likely due to missed conditions, may be fixable with traditional automated program repair techniques.
\end{finding}

\section{Discussion and Future Work}
\label{sec:discussion}

{

Our findings shed light on several future directions to improve the quality and reliability of LLM-generated code. 

\smalltitle{Repairing Errors in LLM-generated Code.} Our Finding~\ref{finding:1} and Finding~\ref{finding:3} reveal that errors made by different LLMs exhibit a wide range of different semantic and syntactic characteristics. Single-line logic errors such as \textit{if errors} and \textit{incorrect function arguments} may be relatively easy to fix, since existing automated program repair (APR) techniques are specialized to fix such errors~\cite{xiong2017precise, mechtaev2016angelix, ghanbari2019practical, jiang2018shaping, liu2019tbar, li2022dear, ye2022selfapr}. However, our Finding~\ref{finding:3} indicates that many errors made by code LLMs involve multiple lines of code, which require substantial edits to fix.

To overcome the limitations of traditional APR techniques, recent work has proposed to leverage LLMs to build program repair agents that are capable of synthesizing more complex patches~\cite{fan2023automated, zhang2024autocoderover, xia2024agentless}. These approaches typically rely on \textit{coarse-grained} information such as error messages to guide the repair process. By contrast, our taxonomy provides a more detailed analysis of code generation errors based on their semantic and syntactic characteristics. We believe that leveraging this fine-grained information to repair errors in LLM-generated code is a promising direction. Specifically, one can first train a machine learning model to predict the types of errors. 
Then, the predicted errors can then be encoded in the prompt to guide the repair agent. For instance, instead of prompting the repair agent with ``{\em fix the bug at Line 6},'' one can prompt the agent with ``{\em fix the incorrect function arguments at Line 6},'' which may lead to a more accurate patch.

{\responseref{}

To demonstrate the effectiveness of repairing code generation errors with our labeled characteristics, we compared the performance of prompting GPT-4 to repair its incorrect code solutions with and without labeled semantic and syntactic characteristics (detailed prompts are available in our repository~\cite{ourwebsite}
). Since GPT-4 generated only 18 incorrect solutions on the HumanEval dataset, we included incorrect solutions generated by GPT-4 from an additional dataset, BigCodeBench-Hard~\cite{zhuo2024bigcodebench}, to increase the sample size. We selected BigCodeBench-Hard since it provides more challenging coding tasks compared to HumanEval and can better demonstrate the potential of fixing code generation errors using our taxonomy. In total, our sample set includes 18 incorrect solutions from HumanEval and 84 incorrect solutions from BigCodeBench-Hard. All these solutions were generated using the original prompt provided by each dataset and a decoding temperature of 0, which is consistent with the settings in Sec.~\ref{sec:collection}. Two authors independently labeled the error characteristics of these solutions using our taxonomy. The Fleiss’ Kappa scores for semantic and syntactic characteristics were 0.91 and 0.93, indicating almost perfect agreement. All disagreements were resolved through discussion. Finally, we prompted
GPT-4 to repair these incorrect solutions using the labeled error characteristics. 

On HumanEval, GPT-4 was able to repair 7 out of 18 incorrect solutions by prompting with our labeled error characteristics, while it repaired only 4 out of 18 incorrect solutions without these characteristics. On the more challenging dataset, BigCodeBench-Hard, our method shows a bigger improvement over the baseline. Specifically, GPT-4 was able to repair 12 incorrect solutions with our labeled error characteristics while only 3 incorrect solutions without these characteristics. These results demonstrated that our taxonomy and labeled error characteristics are even more helpful when repairing incorrect solutions for challenging tasks. 

}

\smalltitle{Fault Localization for LLM-generated Code.} Precisely locating the error location is an important first step for fixing code generation errors. As shown by Finding~\ref{finding:3}, code generation errors can occur in a variety of code constructs, which poses challenges to locating them precisely. Although many fault localization approaches have been proposed~\cite{wen2019historical, abreu2009spectrum, li2019deepfl, li2021fault}, these methods may still fall short in locating errors in LLM-generated code. Traditional approaches, such as spectrum-based fault localization, rely heavily on high-quality test cases that comprehensively examine different execution paths in a program. However, it is effortful and time-consuming to design test cases, which limits the utility of these approaches. Recently, deep learning has also been applied to fault localization~\cite{li2019deepfl, li2021fault}, typically using features extracted from source code and error messages. For code generated by LLMs, there is richer information from different sources, such as logits, token probability distribution, and self-attention scores at each decoding step. It would be interesting to investigate whether such information, together with error messages and other code characteristics, can be leveraged to predict at which step the model may start generating incorrect code.  

\smalltitle{Estimating the Correctness of LLM-generated Code.} Our Finding~\ref{finding:long_prompt_garbage} shows that current LLMs continue to struggle with understanding and solving complex task requirements. This observation raises several interesting research questions. First, for a given code generation LLM, can we develop a method to estimate the upper bound of task complexity that this LLM can handle? In this study, we have only experimented with simple metrics such as prompt length and LOC of the correct solution. Further investigation on this question could help assess the capabilities and limitations of a code generation LLM more accurately, leading to more reliable and trustworthy code generation. Second, for a given programming task, can we use prompt characteristics, such as length, to predict the task’s difficulty and the correctness of the generated code? Some recent work leverages LLMs as a judge for code correctness without the need of test cases~\cite{zhuo2024ice, tong2024codejudge}. While they achieve promising accuracy on simple benchmarks (e.g., 73.13\% by CodeJudge~\cite{tong2024codejudge} on HumanEval), the accuracy on more challenging ones needs more improvement (e.g., 54.56\% on BigCodeBench). Such techniques could help developers better allocate their code review and testing effort based on code correctness estimation.

}

\section{Related Work}



\smalltitle{LLM-based Code Generation.} LLMs have shown great potentials in code-related tasks, such as generation~\cite{fried2022incoder, nijkamp2022codegen, wang-etal-2023-codet5, zhu2024grammart5}, summarization~\cite{ahmad2020transformer, niu2022spt, cai2024fly}, 
{understanding~\cite{wang2022bridging, ding2024traced}}, 
{search~\cite{feng2020codebert, zhang2022diet, jain-etal-2023-contraclm}}, 
and translation~\cite{roziere2020unsupervised}. Recent work on code generation can then be roughly categorized into three groups: (1) developing high-quality training data~\cite{lozhkov2024starcoder, guo2024deepseek, ding2024semcoder}, (2) developing better instruction fine-tuning techniques~\cite{roziere2023code, luo2024wizardcoder, wei2023magicoder}, and (3) developing better prompting strategies~\cite{zhang-etal-2023-self, bairi2023codeplan, gao2023makes, chen2023codet, olausson2023self, li2023skcoder, chen2023teaching, ding2024cycle, madaan2024self}.
Compared to these efforts, our research focuses on analyzing the errors {LLMs} produce and deriving empirical insights to help develop new methods in the direction and contribute to better LLM-enabled intelligent software engineering.

\smalltitle{Quality of LLM-generated Code.} Evaluating the quality of LLM-generated code can reveal the current approach’s shortcomings and guide future improvement. While prior research has delved into facets such as robustness~\cite{ arakelyan2023exploring, liu2023reliability}, security~\cite{pearce2022asleep, steenhoek2024comprehensive}, and attention alignment~\cite{kou2023model}, studies focusing on the correctness of LLM-generated code~\cite{chen2021evaluating, nguyen2022empirical, jesse2023large, liu2023refining, liu2023no} are of particular relevance to our work.


Liu~\etal~\cite{liu2023refining} evaluated the quality of code generated by ChatGPT, addressing factors such as compilation/runtime errors and coding style. Different from their work, we investigated a broader range of LLMs and analyzed  fine-grained error characteristics and their correlations with task complexity, test pass rates, etc. Pan \etal~\cite{pan2023understanding} introduced a taxonomy for LLM's code translation bugs. A key distinction is in the origin of errors, where {Pan \etal's work focuses on code-to-code translation, while our work focuses on NL-to-code generation.}
Our study also identifies distinct semantic error characteristics (e.g., \textit{wrong (logical) direction}) and syntactic error  characteristics (e.g., \textit{missing/incorrect code block}).

Finally, Liu \etal~\cite{liu2023no} conducted a study of the quality of code generated by ChatGPT, assessing their correctness, understandability, and security. In their examination of code correctness, {they} primarily focused on compile errors {and runtime crashes}. Our research differs from theirs in two key respects: (1) Our {study} subjects are more diverse with both open-source and closed-source models; 
(2) our taxonomy also considers behavior deviations informed by test failures in addition to compiling errors and crashes. Based on the findings, we further provide actionable suggestions and implications for future enhancements in LLM-enabled code generation.



\smalltitle{Taxonomy on Software Defects.} Building a systematic taxonomy of software bugs can help stakeholders understand the pitfalls of target systems and provide guidance for better development practices. One of the early attempts was the orthogonal defect classification proposed by IBM Research~\cite{chillarege1991defect, chillarege1992orthogonal}. 
Since then, numerous endeavors have been made to construct defect taxonomies targeting different programming languages~\cite{le2015manybugs, tan2017codeflaws, hanam2016discovering, gyimesi2019bugsjs, pan2009toward, zhang2022excepy, rahman2023come, lin2017quixbugs, kochhar2016large}
or different applications~\cite{rahman2020gang, humbatova2020taxonomy}. 
Unlike previous attempts that focused on bugs in real-world software projects, our work focuses on LLM-generated code. Our findings reveal that a large portion of code errors made by LLMs exhibit complex semantic characteristics rather than subtle errors usually introduced by human programmers.

\section{Threats to Validity}
\label{sec:threats}
\smalltitle{Threats to Internal Validity.} One potential threat comes from our manual analysis process, where labelers may have different opinions and sometimes may make mistakes. To mitigate this, four of the authors first performed open coding and iteratively refined our codebook until a substantial agreement was achieved before two of the authors labeled the complete dataset. Our final Fleiss' Kappa regarding the semantic and syntactic characteristics are 0.91 and 0.91, respectively, indicating almost perfect agreement~\cite{landis1977measurement}.

\smalltitle{Threats to External Validity.} One potential threat lies in the choice of dataset. Considering the extensive labeling effort (e.g., running the code and performing step-by-step debugging to locate the root cause), we have only labeled one dataset. In future work, one may consider labeling more datasets to confirm the generalizability of our findings. Nevertheless, given the size of our labeled dataset (557 code snippets), we believe our findings can be generalized to other similar datasets~\cite{austin2021program, hendrycksapps2021, zhuo2024bigcodebench}. \responseline{Additionally, our study has only explored errors from function-level code generation tasks. There are many other code generation tasks (e.g., class-level code generation~\cite{du2023classeval}), as well as other code-related tasks such as code summarization and refactoring. In future work, it is interesting to investigate whether these tasks share similar error characteristics with function-level code generation errors.}

Moreover, our study only covers Python programs, which might not generalize to programming languages that are very different from Python, such as PHP and Rust. We chose Python because it is one of the most popular object-oriented programming languages. In future work, we plan to expand our study with programming tasks from other languages.

Finally, we have only experimented with six LLMs \responseline{and one prompting strategy} in this study due to the intensive labeling effort (i.e., 328 person-hours for labeling 557 incorrect code snippets). Given that code generation with LLMs is a fast-developing research field, we plan to expand our study by labeling errors from more recent code LLMs, e.g., MagiCoder~\cite{wei2023magicoder},  CodeLlama~\cite{roziere2023code}, etc. \responseline{Additionally, our findings may not generalize to code generation errors with different prompting strategies, e.g., chain-of-thought~\cite{wei2022chain}, self-debugging~\cite{chen2023teaching}, etc. It is worthwhile to investigate code generation errors with more advanced prompting strategies in future work.} 

\smalltitle{Threats to Construct Validity.} In RQ3, we use the length of the task description and the length of the correct solution to estimate the task complexity. Although they may not be the best metrics for such estimation, we believe this is a pragmatic choice since there is no commonly used metric for measuring task complexity for LLM's code generation. In future work, researchers may consider designing new metrics to estimate the task complexity and investigate its correlation with LLM's code generation capabilities.

\section{Conclusion}
This paper presents an empirical study on code generation errors made by LLMs. We first derived a taxonomy of LLMs' code generation errors based on six popular LLMs' failure cases on the HumanEval dataset~\cite{chen2021evaluating} through open coding and thematic analysis. We labeled a total of 557 errors committed by these LLMs according to the taxonomy. We found that these LLMs exhibited different distributions of semantic and syntactic error characteristics. We further analyzed the bug-fixing effort,
the impact of task complexity, and the correlation between test pass rates and different kinds of errors. In the end, we discussed the implications of our study and propose future research opportunities for improving the quality and reliability of code LLMs.

\section*{Acknowledgment}
This work was supported in part by Canada CIFAR AI Chairs Program, Natural Sciences and Engineering Research Council of Canada, JST CRONOS Grant (No.JPMJCS24K8), JSPS KAKENHI Grant (No.JP21H04877, No.JP23H03372, and No.JP24K02920), and the Autoware Foundation. This work was also supported in part by NSF CCF-2340408.

\bibliographystyle{IEEEtran}
\bibliography{reference}

\end{document}